\documentclass{aa}
\usepackage{epsfig}
\usepackage{times}

\begin{document}
   \thesaurus{11 (11.03.1;11.03.4 \object{Pers-Pisc supercluster};12.03.4;12.12.1) }

   \title{Kinematics of the local universe IX.}
   \subtitle{The Perseus-Pisces supercluster and
        the Tolman-Bondi model} 
   \author{M.O. Hanski\inst{1}, G. Theureau\inst{2,3}, T. Ekholm\inst{4,1} \and
        P. Teerikorpi\inst{1}}
   \offprints{M. Hanski }
   \institute{Tuorla Observatory, 21500 Piikki\"{o}, Finland;
        mihanski, timoek, pekkatee@astro.utu.fi
         \and 
Observatoire de Meudon, 92195 Meudon CEDEX, France; gilles.theureau@obspm.fr
\and
Osservatorio di Capodimonte, Via Moiariello 16, 80131 Naples, Italy
         \and
Observatoire de Lyon, F69561 Saint-Genis Laval CEDEX, France}
   \date{Received / Accepted }

\authorrunning{Hanski, M.O.\ et al.}
\titlerunning{Perseus-Pisces and Tolman-Bondi}
   \maketitle

   \begin{abstract}
The matter distribution around the Perseus-Pisces (PP) supercluster is studied
by comparing
peculiar velocities given by the Kinematics of the local universe (KLUN) 
galaxy sample to those predicted by Tolman-Bondi (TB)
models. To restrict the TB solutions we first solve the mass
of the densest part of PP. This part is
identified as a sphere at $(l,b)=(140.2\degr , 
-22.0\degr )$, $d \approx 50 \, h^{-1}$~Mpc  having a radius of
$15 \, h^{-1}$~Mpc.  This sphere surrounds the main part of the PP ridge
and four most prominent clusters of the region. Using virial-like
mass estimators we calculate the cluster masses and obtain the
upper and lower limits for the mass inside the $15 \, h^{-1}$~Mpc sphere:
$M_{\rm PP} = 4$ -- $7\, h^{-1}10^{15} M_\odot$.
This corresponds to a mass overdensity $\delta_{\rm PP} \approx 4$, or
$\rho_{\rm PP}= 1$  -- $2\, 
\rho_{\rm cr}$. Mass to light ratios of the clusters are  
$M/L =$ 200 -- 600 $h\, M_\odot / L_\odot$, giving mass density ratio
$\Omega_0 = 0.1$ -- $0.3$,
if the value of $M/L$ is assumed to be representative elsewhere in the 
universe.
 
We estimate a radial density distribution around the PP core using 
two toy models and a smoothed density distribution
observed for IRAS galaxies.
The cosmological density parameters $(\Omega_0,
\Omega_\Lambda )$ and the PP mass are free
parameters in the TB calculations. The KLUN velocities, obtained 
by Tully-Fisher relation and the normalized distance method, are adjusted
by the Local Group (LG) infall
velocity towards PP.
Comparison of the TB velocities to KLUN data
points indicates that the infall velocity $v_{\rm inf} < 100$~km~s$^{-1}$. 
Allowing $M_{\rm PP}$ to vary within the limits given above we get constraints
for the value of $\Omega_0$; $\Omega_0=0.2$ -- $0.4$ are prefered to the more
extreme values, $\Omega_0=0.1$ or $1$. A choice of either $\Omega_\Lambda=
1-\Omega_0$ or $0$ do not cause any significant changes in the results.

The validity of the TB model in complex environments
is studied with an N-body simulation. There we see that the radially
averaged velocity fields around simulated clusters are compatible with
the corresponding TB velocities. 
This confirms the applicability of the TB model around large galaxy
concentrations,
providing that smoothed density and radially averaged velocity 
fields are used.


\end{abstract}

\section{Introduction \label{sec:int}}

Tolman-Bondi (TB) model is the analytical solution of a spherically symmetric
mass distribution in an expanding universe (Tolman \cite{tolm34}, Bondi
\cite{bond47}). It has been succesfully applied
to the Virgo system by Teerikorpi et al.\ (\cite{teer92}) and, recently,
by Ekholm et al.\ (\cite{ekho99}) for describing
the peculiar velocity field and estimating the
total mass of the system, even though the galaxy distribution in Virgo
deviates notably from spherical. 
Motivation for the use of the TB model
is that it is {\em simple} and {\em analytical}. 
It is a first step towards approximating an
inhomogeneous universe. Bondi (\cite{bond47}) himself wrote that
``the assumption of spherical symmetry supplies us with a model which
lies between the completely homogeneous models of cosmology and the actual
universe with its irregularities''. 

TB results are valid
with appropriate smoothing; the observed
velocity field must be averaged over 
concentric shells around a galaxy 
concentration, when compared to the TB results.
Thus the TB model does not give the results in as much details
as the various more commonly used gravitational instability -- reconstruction
methods. On the other hand, it does not
require any assumptions for the breakdown of the theory in the non-linear
region.
 
In order to test whether the TB model may be applied to 
an apparently still less favourable situation than Virgo, we study the 
Perseus-Pisces supercluster, where a TB-like deviation in the Hubble flow
has already been observed (Theureau et al.\ \cite{theuA98}).
Our criterion for the success of the TB method is whether the mass inferred
using TB is comparable to virial and luminosity $\times$
$M/L$ estimate for the region, and whether the peculiar velocity field
is adequately described by the model.  

The local $d<70 h^{-1}$ Mpc\footnote{$h=H_0/$(100 km~s$^{-1}$~Mpc$^{-1}$)}
universe is dominated by two large mass concentrations, the Great Attractor
and the Perseus-Pisces (PP) complexes. According to Sigad et al.\ 
(\cite{siga98}) and Dekel et al.\ (\cite{deke99}) both regions have 
overdensities $\delta=1$ -- $1.5$,
where $\delta \equiv [\rho(x) - \bar{\rho}]/\bar{\rho}$,
$\rho(x)$ is the local, and $\bar{\rho}$ the average density. Both studies
used a Gaussian smoothing of 12 $h^{-1}$ Mpc and assumed $\Omega_0=1$. Thus the
velocity field surrounding PP should reveal a strong 
infall towards the supercluster's center of mass.
For a galaxy at around half way from PP to us the infall velocity
should be $\approx 500$ km~s$^{-1}$.
However, the infall signature is difficult to reveal because of 
large uncertainties in redshift independent distance measurements,
e.g.\ those provided by Tully-Fisher (TF) or Fundamental Plane (FP) methods.
Uncertainties are caused by  statistical biases difficult to
correct for in a heterogeneous environment. Additional inaccuracies are
raised by the filamentary
shape of the structure, consisting of an alignment of
several subclusters, which complicates considerably
any attempt to model the velocity field.

\subsection{Previous studies of Perseus-Pisces}

In the past ten years PP has been investigated by numerous authors
using various distance criteria and statistical methods. The region is 
now well covered by great amount of data.
From a large reshift sample of 3311 galaxies in the region bounded by
22$^h <$ R.A. $ < 4^h$, +20$^{\circ} <$ Dec. $< +50^{\circ}$,
Wegner et al.\ (\cite{wegn93}) described the Perseus-Pisces complex
as follows: The supercluster 
consists of a continuous arrangement of high density clusters and groups, 
among which the \object{Pisces}, \object{A262}, and \object{Perseus} 
(\object{A426}) clusters
are the most prominent features.  Its main ridge extends at least
50 $h^{-1}$ Mpc from Pegasus eastward to the Perseus cluster, with a width in 
the plane of the sky of about 5--10 $h^{-1}$ Mpc, and a typical depth in 
redshift 
of 250--500 km~s$^{-1}$. All the objects in the ridge are lying roughly 
at the  same distance from us, at $\sim$ 5000 km~s$^{-1}$.

First kinematical studies (Willick \cite{will90}, \cite{will91},
Courteau et al.\  \cite{cour93}) based on an r-band TF sample of 
355 galaxies, suggested
that the whole structure takes part on a large scale flow defined
by the Great Attractor (see e.g.\ Lynden-Bell et al.\ \cite{lynd88}, Dekel 
\cite{deke94}), with a mean velocity of $\sim$ $350$ km~s$^{-1}$
towards the Local Group (LG). Han \& Mould (\cite{hanm92}) confirmed 
this result independently  using a whole sky TF sample of 21 clusters. 
Hudson et al.\ (\cite{huds97}) combined their inverse FP relation measurements 
of seven clusters in the PP region with data of nine other clusters 
from the literature and concluded that the whole region has a mean 
CMB frame bulk motion of $420 \pm 280$
km~s$^{-1}$ towards galactic coordinates $(l,b)=(262\degr ,-25\degr )$.
POTENT reconstruction method of the density and velocity field for
Mark III galaxies (Dekel et al.\ \cite{deke99}) gave different results.
There it was found that the center of PP is roughly at rest in the LG frame
and the regions between the LG and PP are infalling towards PP. 

Most studies of the velocity field of galaxies suggest
that the kinematics of the local 
universe can be described by a model in which galaxies are infalling to two
mass concentrations, one in PP and one in Hydra-Centaurus (e.g.\ Han \&
Mould \cite{hanm92}, Dekel et al.\ \cite{deke99}). 
However, 
Freudling et al.\ (\cite{freu95}) and da Costa et al.\ (\cite{daco96}) 
used a method of analysis, based on Monte-Carlo simulations, in order to
control biases and gain a better sampling of the sky. Their studies 
revealed a clear infall motion towards PP, while the GA was found less dominant
than originally believed. 

The differences in the results stated above
may be caused by uncertainties in the reconstruction method,
in the corrections for the selection effects, and/or in the distance
indicator calibration. For example, an error in the TF zero point 
would cause a systematic error that resembles a Hubble flow
in the peculiar velocity field. Dekel et al.\ (\cite{deke99}) also 
note that the results in the PP region are more sensitive to noise than
the other regions.

\subsection{What is new in our approach to PP?}

In a previous study using the KLUN galaxies,
a full infall pattern, 
both in the front and the backside of the PP supercluster was shown, for 
the first time.  A new Malmquist
bias correction method for the TF distance data was used (Theureau et al.\ 
\cite{theuA98}).
In the present paper we investigate the dynamics of galaxies in the 
PP surroundings in more detail. We study the spatial distribution 
and  peculiar velocities of the galaxies
using the Malmquist corrected TF distances.
The observed velocity field is averaged over spherical shells
around the PP center and compared
to the results given  by TB calculations. 

The outline of the paper is
as follows: In Sect.\ \ref{sec:meth} we briefly present 
the data and summarize the normalized distance method (NDM) 
that has been used for deriving unbiased average distances and 
peculiar velocities.
In Sect.\ \ref{sec:vir} we calculate the virial masses of 
the most prominent clusters
in the PP region and estimate the mass of the whole central region of 
PP from these masses.
Section \ref{sec:tb} includes a description of the Tolman-Bondi model and
its application to PP. The model is also tested on an N-body 
simulation of galaxy clusters (Sect.\ \ref{sec:sim}). 
In Sect.\ \ref{sec:comp} we compare the TB model
to the infall pattern observed with the KLUN data.
Finally, the main results are summarized in 
Sect.\ \ref{sec:res}.

\section{Towards unbiased peculiar velocities \label{sec:meth}}

The following notation is appointed for different distance and
velocity variables:
\begin{itemize}
\item{$d$ for distances from the LG}
\item{$r$ for distances from the PP center}
\item{capital $R$ for fixed distances: $R_{\rm PP} (=50 h^{-1}$ Mpc) 
is the distance between the LG and the PP center, $R_{\rm s} (=15 h^{-1}$ Mpc) 
is the radius of the main region of the PP concentration}
\item{small $v$ for peculiar velocities, $v_r$ is the peculiar velocity
with respect to the PP center, $v_d$ is the peculiar velocity with respect
to the LG}
\item{capital $V$ for observed (Hubble flow + peculiar) velocities,
corrected for the infall towards Virgo as described in Sect.\ \ref{sect:data}}
\end{itemize}
Exceptions are
\begin{itemize}
\item quantities with subindex 25: $D_{25}$ is the standard 
notation for the isophotal $B=25$ mag diameter of a galaxy 
and $R_{25}$ is the corresponding axis ratio
\item $V_{\rm m}$ is the maximum of the rotational velocity of a galaxy
\end{itemize}

For the Hubble constant we assume $H_0 =60$ km~s$^{-1}$~Mpc$^{-1}$, 
or indicate it by the term 
$h \equiv H_0/$(100~km~s$^{-1}$~Mpc$^{-1}$). The value is needed in 
calculating the  radial part of the peculiar velocities,
\begin{equation}
v_{d,{\rm obs}} = V - H_0 d,
\label{eq:vpec}
\end{equation}
where the subscript ``obs'' emphasizes that only this component of the 
peculiar velocity can be directly observed.

The main difficulties in velocity field studies
are caused by the large uncertainties in distance 
determination and by the resulting biases.
Attempts to overcome these problems require a few things. 
Firstly, one must know thoroughly the 
sample, how it was collected,
and to which extent it can be assumed complete.
Secondly, systematic effects in the distance criterion have to be 
modeled, so that one can derive for a given application an
unbiased set of distances. In particular, biases will be different if one
estimates peculiar velocities at a given derived distance, at a given
true distance, or at a given observed radial velocity. 
Thirdly, the sample has to be deep enough and should cover a sufficiently 
large solid angle in the sky for the average relations, needed for the
bias correction, to be free from
local perturbations in the density or the velocity field. 
This last requirement is similar as needed for a satisfactory determination
of the Hubble constant.

\subsection{The data \label{sect:data}}

We use the KLUN (Kinematics of the local universe) sample 
consisting of 6600 spiral galaxies
having measurements of isophotal diameter $D_{25}$, HI line width, radial
velocity, and B-magnitude. This sample has been
fully described in previous papers of this series, e.g.\ 
by Theureau et al.\ (\cite{theu97}), so we only summarize 
here the main characteristics. 

KLUN is selected according to apparent diameter, and is complete down to 
$D_{25}=1\farcm 6$  (Paturel et al.\ 
\cite{patu94}), covering the type range Sa--Sdm ($T$=1--8). 
The data are extracted from LEDA (Lyon-Meudon Extragalactic Database,
Paturel et al.\ \cite{patu97}) and
complemented with optical redshifts measured with ESO and OHP telescopes, 
and  HI spectra obtained with Nan\c{c}ay and Parkes radio telescopes 
(Bottinelli et al.\ \cite{bott92}, \cite{bott93}, di Nella et al.\ 
\cite{dine96}, Theureau et al.\ \cite{theuB98}). According to LEDA 
precepts, all astrophysical parameters have been homogenized and
reduced to a standard and common system (Paturel et al.\ \cite{patu97}). 
Isophotal $D_{25}$ diameters and apparent B-magnitudes are corrected
for galactic extinction according to Fouqu\'{e} \& Paturel (\cite{fouq85}), 
and for inclination, i.e.\ the opacity effect, as explained in Bottinelli
et al.\ (\cite{bott95}). HI line widths, reduced to the standard levels of 20\%
and 50\%, are corrected for internal velocity dispersion according to
Tully \& Fouqu\'{e} (\cite{tull85}). Heliocentric radial velocities 
are changed to the rest frame of the LG centroid   
following Yahil et al.\ (\cite{yahi77}) and
corrected for a Virgo-centric component,
assuming an infall of the LG towards the Virgo cluster 
$v_0 = 200$~km~s$^{-1}$ (Theureau et al.\ \cite{theu97}, 
Jerjen \& Tammann \cite{jerj93}) 
and a radial velocity of Virgo $V_{0,{\rm Vir}} = 980$ km~s$^{-1}$ 
(Mould et al.\ \cite{moul80}), within a linear infall model (Peebles 
\cite{peeb76}).

In the TF relation studies we need to make some restrictions to the data.
Galaxies close to the Galactic plane ($|b| \leq 15^{\circ}$) are excluded
due to the large uncertainties in the Galactic extinction correction.
Face-on galaxies, $\log{R_{25}} < 0.07$, 
are excluded to prevent large errors on calculation of the 
rotational velocity parameter $p \equiv \log{V_{\rm m}}$, where $V_{\rm m}$
is the maximal rotational velocity of a galaxy, measured from the HI line 
width. 

Absolute TF distances are obtained using the 
calibration given by Theureau (\cite{theu98}).   TF 
scatter was there reduced by 30\% by taking into account the dependence 
of the TF zero point on mean surface brightnesses of the galaxies. 
The NDM first order Malmquist corrections are calculated on the
basis of the whole sky sample, according to Theureau et al.\ (\cite{theuA98}).
The sample with thus corrected TF distances, 
restricted to the strictly diameter
complete part, contains finally 2800 spiral galaxies.

\subsection{The normalized distance method (NDM) \label{sec:ndm}}

The concept of normalized distances was introduced in a theoretical
discussion by Teerikorpi (\cite{teer84}).
The method was first applied 
by  Bottinelli et al.\ (\cite{bott86}) and improved 
in a series of papers, most recently
by Theureau et al.\ (\cite{theu97}). A theoretical description in terms of 
probability densities was given by Theureau et al.\ (\cite{theuA98}),
where the authors showed that the method can be extended to make
Malmquist bias corrections to galaxy distances even beyond the so-called
unbiased plateau. The unbiased plateau has been profitably used
for determination of the unbiased value of $H_0$ and has the advantage
that one does not need model-dependent bias corrections.
\begin{figure}
\epsfig{file=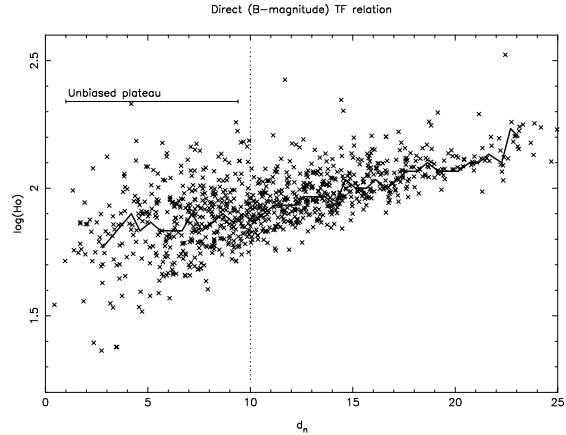,angle=270,width=8cm}
\caption{The $\log H_0 $ vs.\ $d_{\rm n}$ diagram of KLUN galaxies.
The line, indicating the average $\langle \log H_0 \rangle(d_{\rm n})$,
is (approximately) horizontal in the unbiased region at small $ d_{\rm n}$.
Malmquist bias makes the curve to rise at larger $ d_{\rm n}$.}
\label{fig:h0dn}
\end{figure}

In order to understand how the NDM works,
let us define classes of galaxies as follows:\ objects belong to a given
class if they have identical values of log rotational velocity 
$p$, Galactic extinction correction 
$a_{\rm g}$, inclination $R_{25}$, morphological type $T$ 
and mean surface brightness $\Sigma$.
The main idea of the NDM is that for a given magnitude or 
diameter limited sample, and for a given TF scatter the 
behaviour of the Malmquist bias\footnote{In fact the Malmquist bias
of the 2nd kind, as termed by Teerikorpi (\cite{teer97}).} 
as a function of true distance is similar for 
all galaxies belonging to the same class. The class is 
assumed to have a Gaussian
distribution of absolute magnitudes. Then, for each class, there
is a unique curve of evolution of the bias, which is conveniently expressed as
the curve of observed average value of the Hubble parameter 
$\langle \log{H_0} \rangle=\langle \log{V_{\rm c}} - \log{d_{\rm TF}} \rangle$ 
vs.\ distance ($d$), where $d_{\rm TF}$ is the distance given by TF relation,
\begin{equation}
\log{d_{\rm TF}} = a_{\rm TF} p + b_{\rm TF}(\Sigma)
-\log{D_{25}}-\log{\frac{\pi}{108}};
\end{equation}
$a_{\rm TF}$ is the TF slope, $b_{\rm TF}(\Sigma)$ the TF zero point,
dependent on the surface brightness $\Sigma$ of the galaxy, and $\frac{\pi}{108}$
is a constant for converting the units (linear TF-diameter in kpc and
$D_{25}$ in $0\farcm 1$ to $d_{\rm TF}$ in Mpc).
 
\begin{figure*}
\epsfig{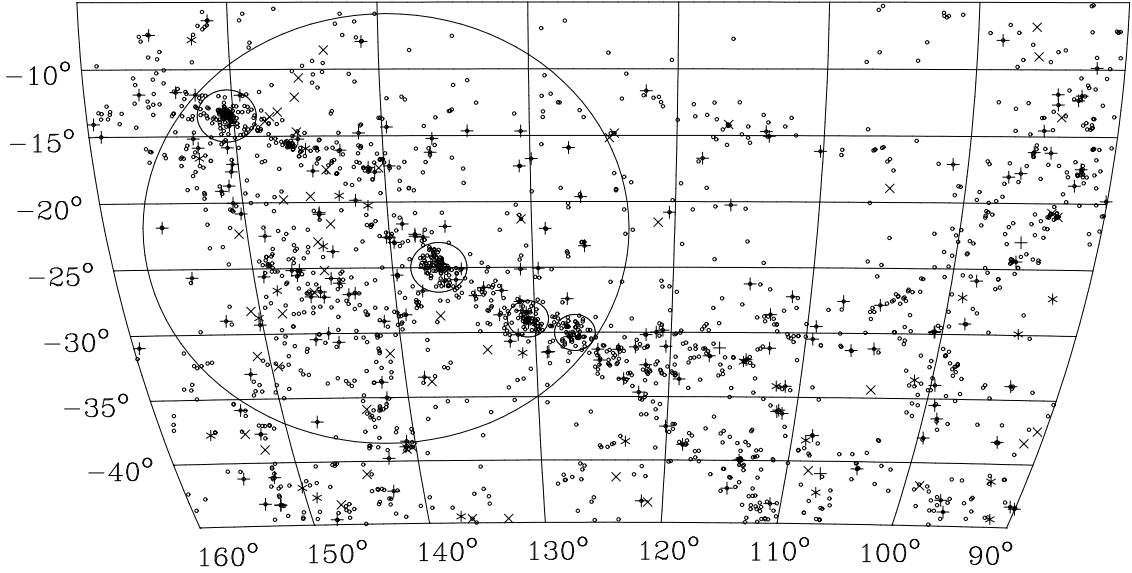}
\caption{LEDA (o) and KLUN galaxies ($+$, $*$ and $\times$) in the PP region.
Large circle is the 15~$h^{-1}$~Mpc sphere at $(140.2\degr ,-22.0\degr )$ 
containing the main concentration. Small circles are the four densest 
clusters, \object{Perseus}, \object{A262}, \object{0122+3305} and 
\object{Pisces}, from up left to down right.}
\label{fig:map}
\end{figure*}

Essentially, the normalized distance technique 
means superimposing all the bias curves
$\langle \log H_0 \rangle(d)$ of the different classes of objects 
by using an appropriate scaling on the distance:
\begin{equation} 
d_{\rm n} = d \cdot f(p,a_{\rm g},R_{25},T,\Sigma) 
\end{equation}
 As a result, the amplitude
of the bias can be expressed merely as a function of the
normalized distance ($d_{\rm n}$), the apparent limit of the sample 
($m_{\rm lim}$ or $D_{\rm 25,lim}$), and the TF scatter ($\sigma_{\rm TF}$).
Figure \ref{fig:h0dn} shows the $\langle \log H_0 \rangle $ 
vs.\ $d_{\rm n}$ diagram of KLUN galaxies in the case of the B magnitude
TF relation. The unbiased part
of the sample is identified as a plateau of $\langle \log H_0 \rangle =$ 
constantat, at low values of $d_{\rm n}$.
Note, however, that for calculation of $d_{\rm n}$, one needs the true
distance $d$, usually derived as the so-called kinematical distance, e.g.\
in the case of the Hubble law $d=d_{\rm kin}\equiv V/H_0$.

The average bias correction is valid only if the following assumptions are 
fulfilled:
\begin{itemize}
\item [-]{\it H1} : the sample is complete and strictly diameter (magnitude)
limited 
\item [-]{\it H2} : TF residuals follow a Gaussian distribution, and the 
TF scatter does not depend on the value of $p$
\item [-]{\it H3} : the selection in apparent diameter (magnitude)
is independent of the selections in redshift and $p$
\item [-]{\it H4} : the radial velocity field integrated over all the 
directions vanishes, i.e.\ except the Hubble flow, there is no coherent 
flow at the scale of the whole sample
\end{itemize}

Assumption {\it H1} is obtained by applying a strict cut-off 
at the completeness limit of the sample. {\it H2} is in agreement 
with what is observed in the case of the B-band TF relation 
(Theureau et al.\, \cite{theu97}). {\it H3} might be difficult to prove,
if observational errors on apparent magnitudes or diameters are correlated with
the errors on $p\,$:\ both are linked to the inclination of the
galaxy through the internal extinction correction and the de-projection of
the observed HI line width. However, this effect is insignificant in the 
case of the diameter TF relation.  

As for assumption {\it H4},
a list of recent measurements of large-scale bulk motions is
given by Strauss (\cite{stra00}). The list is somewhat divergent; roughly
half of the surveys find no bulk flow at scales $V\geq$ 6000 km~s$^{-1}$,
the rest claim bulk flows of $300$--$600$ km~s$^{-1}$. We will return to
the question of large-scale flow within the KLUN sample in the future,
for now we assume that assumption {\it H4} is true for our data.

Finally, the corrected distance is given by 
\begin{equation}
\log{d_{\rm c}} = \log{d_{\rm TF}} + \sqrt{\frac{2}{\pi}} \cdot \frac{ 
\sigma_{\rm TF}
\exp{\left( -\frac{(\omega -\log{D_{\rm 25,lim}})^2}{2 \sigma_{\rm TF}^2}
\right)}}{1-\mbox{erf}\left( \frac{\omega -\log{D_{\rm 25,lim}}}{\sqrt{2}
\sigma_{\rm TF}} \right)},
\end{equation}
(Theureau et al.\ \cite{theuA98}), 
where the (log) normalized distance is marked with $\omega$,
\begin{eqnarray*}
\omega \equiv \log{d_{\rm n}} =\log{d} + a_{\rm TF}(2.7 - p) + 
0.094 a_{\rm g} \\ 
- C \log{R_{25}} - b_{\rm TF}(\Sigma).
\end{eqnarray*}
Here $d=V/H_0$ and parameter $C$ defines the 
correction for inclination (Bottinelli et.\ al, \cite{bott95}).
For a single galaxy class and usual kinematical distance, 
an analogous formula was 
derived by Teerikorpi (\cite{teer75}). It has also appeared in the
``triple-value correction method'' of Sandage (\cite{sand94}). 

We emphasize that this correction
is statistical in nature. When applied to an individual galaxy, it
gives an accurate {\em average} correction only if the velocity field model
(providing $d_{\rm kin}$ and $d$) is correct. When the velocity field
is distorted, say, in the direction of a supercluster, it still gives a first
order correction, but a complete correction would require an iterative 
procedure (see Sect.\ \ref{sect:ite}). Even the first order 
correction essentially 
improves the derived velocity field, as was shown by Theureau et al.\ 
(\cite{theuA98}) for the PP supercluster, but it is notable that the
peculiar velocities will be somewhat 
suppressed if only a first order correction is done.

\section{The virial mass of Perseus-Pisces\label{sec:vir}}

Figure \ref{fig:map} shows a map of the PP region. 
Circles stand for galaxies from LEDA, having radial velocities
between 3000 and 7000 km~s$^{-1}$ .
This limitation allows us to look at the main part of the PP concentration,
lying at $V \approx 5000$ km~s$^{-1}$, practically
free from any contamination from fore/background objects. LEDA is 
regularly updated for all published data and 
thus contains all the recent PP-surveys.

Symbols other than circles are KLUN galaxies. $+$:s are objects within the
velocity interval stated above, $*$:s have $V>7000$ km~s$^{-1}$ and
$\times$:s $V<3000$ km~s$^{-1}$. Only the KLUN galaxies satisfying the 
completeness limit, 
$D_{25}>1\farcm6$, and other criteria demanded by
our TF-method (see Sect.\ \ref{sec:ndm}), are plotted. 
In total, Fig.\ \ref{fig:map} contains
2016 LEDA galaxies and 266 objects from KLUN. 

\begin{figure}
\epsfig{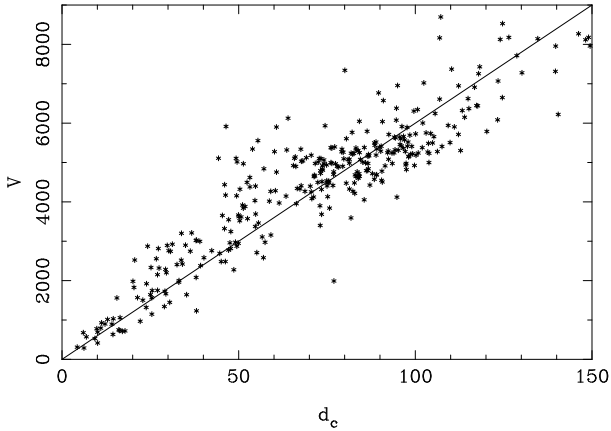}
\caption{The Hubble diagram of KLUN galaxies in the PP region.
$d_{\rm c}$ is the NDM corrected TF distance.}
\label{fig:hub}
\end{figure}  
In Fig.\ \ref{fig:hub} the Hubble diagram of the KLUN PP region 
galaxies is shown. The line corresponds to the Hubble law with
$H_0=60$ km~s$^{-1}$~Mpc$^{-1}$. The infall towards PP is seen
as the surplus of galaxies above the Hubble line at small distances
and below the line at large distances.  

We seek a rough estimate for the amount of matter in the PP region.
First we determine the masses of the densest clusters in the region. 
We define the densest consentrations as the ones having 
the projected galaxy overdensity $\delta_{\rm 2D}>10$.
Here $\delta_{\rm 2D}(x)=(\rho_{\rm 2D}(x)-\bar{\rho}_{\rm 2D})/\bar{
\rho}_{\rm 2D}$, the two dimensional sky densities $\rho_{\rm 2D}(x)$ are
calculated in 1 degree$^2$ regions, and $\bar{\rho}_{\rm 2D}$ 
is the mean value of the whole region shown in Fig.\ \ref{fig:map}.
The limiting factor 10 has no special physical meaning. It was chosen
to give a reasonable number of dense galaxy concentrations.
There are four clusters satisfying this criterium. 
The characteristics of these clusters are listed in Table \ref{tab:1}.
Masses are estimated using
the virial theorem and the alternative methods presented by Heisler
et al.\ (\cite{heis85}), see Table \ref{tab:2} below. 
The effects of possible outsiders and interlopers are tested by
using various radii and redshift intervals in defining cluster members.
Table \ref{tab:2} shows some of the results. For the mass evaluations in 
Table \ref{tab:1}
we adopt $V$-interval of 3000--7000 km~s$^{-1}$, and 
use virial radii of $2^\circ$ for \object{Perseus} and \object{A262}, 
and $1.5^\circ$ for \object{0122+3305} and \object{Pisces}, 
These radii are drawn in Fig.\ \ref{fig:map}.

\begin{table}
\caption[]{Galactic coordinates, numbers of galaxies, mean velocities, virial 
radii in degrees and in $h^{-1}$ Mpc, upper and lower estimates of 
masses in $h^{-1} 10^{14} M_\odot$,
and mass to light ratios in $h M_\odot/L_\odot$, of the four largest
PP clusters in radial velocity interval 3000--7000 km~s$^{-1}$.
Clusters are 1) \object{Perseus}, 2) \object{A262}, 3) \object{0122+3305}, and
4) \object{Pisces}.}
\begin{tabular}{lccrccccc} 
 & $l$ & $b$ & $N$ & $V$ & $\theta_{\rm v}$ & $r_{\rm v}$ &$M$ & 
$ M/L$\\ \hline \hline 
1 & 150.5 & -13.5 & 152 & 5144 & 2.0 & 1.8 & 8--12 & 310--490 \\
2 & 136.7 & -25.0 & 104 & 4878 & 2.0 & 1.7 & 4--6 & 210--370 \\
3 & 130.5 & -29.0 & 67 & 5085 & 1.5 & 1.3 & 4--6 & 260--510 \\
4 & 127.1 & -30.0 & 63 & 5096 & 1.5 & 1.3 & 3--5 &270--550 \\ \hline
\end{tabular}
\label{tab:1}
\end{table}

\begin{table}
\caption[]{Virial, projected, median and average masses by methods of
Heisler et al.\ (\cite{heis85}), and the corresponding $M/L$
ratios  of the four PP clusters. Different radii $\theta_{\rm v}$ are used for
each. The figures indicate that virial and median mass estimators seem
to be less affected by outsiders ($dM/d\theta_{\rm v}$ small) than projected
and average masses. Projected mass is also very sensitive on the coordinates
chosen for the cluster center. This table is calculated for galaxies 
having  $3000$~km~s$^{-1}<V<7000$~km~s$^{-1}$. Changing the $V$ interval
affected all the mass estimators similarly. The limits for $V$  and the 
cluster virial radii are chosen partly by eye, 
simply by using limitations defining a compact galaxy concentration,
and partly by requiring certain level of stability for the obtained
values of mass.
The values adopted in Table \ref{tab:1} are the lowest and the highest 
values from the rows
$\theta_{\rm v}=2.0\degr$ for \object{Perseus} and \object{A262}, 
and $\theta_{\rm v}=1.5\degr$
for \object{0122+3305} and \object{Pisces}.  
Units:\ $[M]= h^{-1} 10^{14}M_\odot$, $[M/L]
= h M_\odot/L_\odot  $ }
\begin{tabular}{crcrrrrc}
$\theta_{\rm v}$ & $N$   & $\langle V \rangle$ & $M_{\rm vi}$&$M_{\rm pr}$ &
 $M_{\rm me}$ & $M_{\rm av}$  & $ M/L $ \\ \hline \hline
\multicolumn{8}{l}{\object{Perseus}} \\ \hline
$ 1.0\degr$ & 115 & 5107 &  5.5 & 6.6 & 4.9 & 5.5 &  244--362 \\
$ 1.5\degr$ & 131 & 5131 &  6.4  &7.9 & 6.1 & 6.9 &  275--393 \\
$ 2.0\degr$ & 152 & 5144 &  8.1  &11.8 &8.1 &10.1 &  309--493 \\
$ 2.5\degr$ & 172 & 5119 &  9.4 & 14.3 &10.0&12.5 &  314--525\\
$ 3.0\degr$ & 183 & 5148 &  10.3 &16.1 &11.2&14.2 &  320--546\\ \hline
\multicolumn{8}{l}{\object{A262}} \\ \hline
$ 1.0\degr$ & 62 & 4965 &  2.8 &  3.4 &  2.3 &  2.7 &  195--342 \\
$ 1.5\degr$ & 93 & 4915 &  3.6 &  4.1 &  2.9 &  3.5 &  168--284 \\
$ 2.0\degr$ & 104& 4878 & 4.3 &  5.6 &  3.8 &  4.6 & 206--369\\
$ 2.5\degr$ & 116& 4892 & 5.1 &  7.0 &  4.7 &  5.6 &  220--405\\
$ 3.0\degr$ & 132&4924 & 5.9 &  8.1 &  5.6 &  6.6 &  233--410 \\ \hline
\multicolumn{8}{l}{\object{0122+3305}} \\ \hline
$ 1.0\degr$ & 49 &5036 &   2.7  & 2.9  & 2.5  & 2.5  & 236--333\\
$ 1.5\degr$ & 67 & 5085 &  4.5  & 5.9  & 3.6  & 4.7  & 260--511\\
$ 2.0\degr$ & 88 & 5063 &  5.6  & 7.1  & 4.5  & 5.8  & 261--502 \\
$ 2.5\degr$& 98 & 5037 & 6.2 &  7.9 &  5.1 &  6.5 &  266--499 \\
$ 3.0\degr$& 125 & 5052 &   7.1&   9.1 &  5.5 &  7.5 & 220--440\\ \hline
 \multicolumn{8}{l}{\object{Pisces}} \\ \hline
$ 1.0\degr$ & 47 & 5073 &   1.9 &  2.9 &  1.7 &  2.1 &  203--409 \\
$ 1.5\degr$ & 63 & 5096 &  2.9 &  4.6 &  2.7 &  3.3 &  270--550\\
$ 2.0\degr$ & 77 & 5105 &  4.2 &  7.1 &  3.7 &  5.1 &  308--713 \\
$ 2.5\degr$ & 95 & 5107 &   5.2 & 8.3 &  4.7 &   6.1 &  329--702\\
$ 3.0\degr$ & 122&5187 &  9.0 & 16.2 &  7.1 & 11.6 &  385--1074\\ \hline 
 
\end{tabular}
\label{tab:2}
\end{table}

\begin{figure}
\epsfig{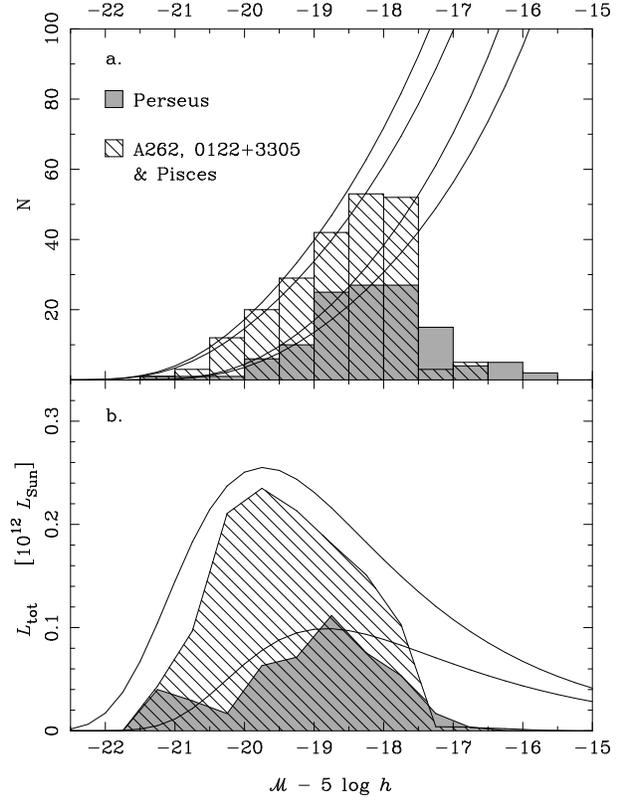}
\caption{Figure a.\ shows the observed absolute magnitude histograms
for the Perseus galaxies (gray bins) and the three other clusters combined
(hatched bins). Two Schechter function curves 
are drawn on both of these histograms. Lower lines
are fitted to the histograms at $\mathcal{M}-5\log h<-18$, 
upper lines are calculated by assuming
completeness of  $80 \%$ for Perseus and
$85 \%$ for the other clusters,  following the latitude dependent
completeness estimate by 
Marinoni et al.\  (\cite{mari99}). Figure b.\ shows 
the corresponding total luminosities in 0.5 magnitude bins. 
The lines are the values predicted by the upper Schechter curves 
of Fig.\ a. By estimating the ratio of
the gray/hatched areas to total areas under the Schechter curves in Fig.\ b.\
we can calculate the amount of luminosity of non-detected galaxies. This is
$42^{+3}_{-2} \%$  for Perseus and $40^{+8}_{-3} \%$ of the total luminosity
for the others.
}
\label{fig:lfC}
\end{figure}

For the $M/L$ ratio derivations in Tables \ref{tab:1} \& \ref{tab:2}
we calculate the total luminosities of the clusters from LEDA
B-band brightnesses and estimate selection effects
using the Schechter function
\begin{equation}
\phi(\mathcal{M}) \propto 10^{0.4(\mathcal{M}^*-\mathcal{M})(\alpha+1)}
\exp{[-10^{0.4(\mathcal{M}^*-\mathcal{M})}]},
\label{eq:lf}
\end{equation}
expressed here in terms of absolute magnitude $\mathcal{M}$. 
The characteristic magnitude, 
$\mathcal{M}^*$, and slope, $\alpha$, of the function are found by fitting  
Eq.\ \ref{eq:lf} to the observed magnitude distribution.
Since the Perseus cluster shows clearly different features from
the other three clusters, the fitting for it is done separately.
Obtained parameters are $\alpha=-1.46^{+0.22}_{-0.26}$ 
and $\mathcal{M}^*-5\log h=-19.6^{+0.4}_{-0.3}$
for \object{Perseus}, and $\alpha=-1.51^{+0.37}_{-0.23}$ 
and $\mathcal{M}^*-5\log h=-20.4^{+0.7}_{-0.7}$ for
the others (Fig.\ \ref{fig:lfC}). The errors are bootstrap $1\, \sigma$
confidence limits. 

Marinoni et al.\ 
(\cite{mari99}) observed for galaxies in large groups
$\alpha=-1.28\pm 0.18$ and $\mathcal{M}^*-5\log h=-20.38 \pm 0.31$.
The difference between our value for \object{Perseus} 
$\mathcal{M}^*$ and the one given by Marinoni et al.\ is probably due to 
environmental differences between \object{Perseus}
and other clusters and/or uncertainties in the brightness measurements
caused by the Galactic absorption. 
Comparison of the total observed luminosities of the clusters to the values
obtained by integrating Eq.\ \ref{eq:lf} indicates that the contribution of
non-detected galaxies is $\mathcal{P}_{\rm nd}=42^{+3}_{-2} \%$ 
for Perseus and $\mathcal{P}_{\rm nd}=40^{+8}_{-3} \%$
for the others, with $1\, \sigma$ bootstrap errors.
The limits for the $M/L$ ratios are then calculated with the upper 
and lower mass values from Table \ref{tab:1} and by estimating
the true total luminosities to be $1/(1-\mathcal{P}_{\rm nd})$,
i.e.\ 1.64--1.80 (Perseus) and 1.60--1.95 (others) $\times$ 
the observed values.  
Resulting $M/L$ ratios are all in the range of 210--550 $h M_\odot/L_\odot$,
in good agreement with other recent estimations of $M/L$ ratios of rich
clusters (e.g.\ Bahcall et al.\ \cite{bahc95}, Carlberg et al.\ 
\cite{carl97}). 

Adding up the masses of the
four clusters in Table \ref{tab:1}, 
we get  $\sim$ 2--3 $h^{-1} 10^{15} 
M_\odot$. The center of mass of PP, using these clusters, is at 
$V_{\rm C}  = 4965$ km~s$^{-1}$, $(l_{\rm C},b_{\rm C})=
(140.2\degr,-22.0\degr)$.   Distances of the clusters   
from this point are 11.5, 3.9, 9.9 and 12.5 $h^{-1}$ Mpc. A sphere at   
$(l_{\rm C},b_{\rm C})$, with  
a radius of $15 h^{-1}$ Mpc surrounds all these main concentrations.  
In addition to the four clusters, this sphere contains 513 other galaxies.  
If we assume that the $M/L$ ratio does not grow in scales larger than  
the rich clusters, as suggested by Bahcall et al.\ (\cite{bahc95}, 
\cite{bahc00}),  
we can then estimate $M_{\rm PP} \approx$ 4--7 $\, h^{-1} 10^{15} M_\odot$, 
where $M_{\rm PP}$ is the mass inside the 15 $h^{-1}$ Mpc sphere.  
The selection effects are assumed to be the same for the galaxies in
the four clusters and for the other galaxies in the $15 h^{-1}$  Mpc sphere.

However, there is some evidence that $M/L$ increases on scales larger than 
clusters. For example Small et al.\ (\cite{smal98}) 
found the mass to light ratio 
to be 564 $h\,M_\odot/L_\odot$ for the Corona Borealis supercluster and
726 $h\,M_\odot/L_\odot$ for a background supercluster. Small et al.\
used the virial and projected mass estimators for the whole bound supercluster.
The Tolman-Bondi 
calculations (next Sect.) show that the PP is also bound, so we
may apply the virial estimators on the PP also. We get
$M_{\rm PP} =$ 2.6, 4.2, 2.8, and 3.3 $\, h^{-1} 10^{15} M_\odot$
for the 836 galaxies in the central $15 h^{-1}$  Mpc sphere in PP,
using the virial, projected, medium, and average mass estimators, 
respectively. 
The corresponding sum of the galaxy luminosities gives 4.4  
$h^{-2} 10^{12} L_\odot$. Taking from above the more conservative limits
for the number of non-detected galaxies, this number is then
increased by a factor 1.60--1.95, giving the total luminosity
$L_{\rm PP} = 7.0$--$8.5\, h^{-2} 10^{12} L_\odot$.
Finally, the mass to light ratio, using these values, is
330--600 $h\,M_\odot/L_\odot$. However, since this mass for
the whole PP core is about equal to the sum of the four clusters
derived above, we doubt the accuracy of the virial and other
estimators in this case, and adopt the values 4--7 $h^{-1} 10^{15} M_\odot$
for $M_{\rm PP}$.

These upper and lower limits for $M_{\rm PP}$ give 
a mean density $\rho_{\rm PP} \approx$ 2--4 $h^2 10^{-26} 
\mbox{ kg~m}^{-3} \approx$ 1--2 $\rho_{\rm cr}$, where $\rho_{\rm cr}$ is
the critical density of the universe. In terms of mass overdensity
this is   $\delta_{\rm PP} \approx$
4, assuming $\Omega_0 \approx$  0.2--0.4.
The lower value of $\Omega_0$ here corresponds naturally to the lower limit 
of $\rho_{\rm PP}$, and high $\Omega_0$ to the higher $\rho_{\rm PP}$. 
The result is in agreement with the value $\delta \approx 1$ ($\Omega_0=1$)
by Sigad et al.\ (\cite{siga98}) and Dekel et al.\ (\cite{deke99}).

\section{Tolman-Bondi method\label{sec:tb}}

\subsection{Theoretical background}

Tolman-Bondi (TB) equations determine the kinematics of a spherically 
symmetric mass concentration in an expanding universe (see e.g.\  
Olson \& Silk \cite{olso79}, and Ekholm \cite{ekho96} for main features of TB, 
and Teerikorpi et al.\  \cite{teer92}, Ekholm \& Teerikorpi \cite{ekho94}, 
and Ekholm et al.\ \cite{ekho99}, \cite{ekho00}
for some recent applications). In practice we study the behaviour of a
spherical layer of matter at distance $r$ from the center of symmetry.
Each of these shells have two dynamical constants. In Newtonian terms 
we can think of these constants as the mass inside the radius of the shell,
\begin{equation} 
M(r)= \int_0^r 4 \pi \rho(r') r'^2 dr'
\end{equation} 
and the energy per mass of the shell,
\begin{equation}
E(r) = \frac{1}{2} \left(\frac{dr}{dt}\right)^2-\frac{GM(r)}{r}-
\frac{\Lambda c^2 r^2}{6}.
\end{equation}
In view of the recent interest in Friedmann models with non-zero
cosmological constant $\Lambda$, we have added its contribution.
The energy term determines whether the mass shells will eventually 
escape ($E(r)>0$) or collapse to the center ($E(r)<0$).
The usual assumption that the mass shells do not overtake each other,
i.e.\ there are no shell crossings,  is followed here.

We attempt to solve the present day velocities of the mass shells, $v_r(r)$,
for the spherically symmetric mass distribution, with the initial condition
that at some point in the past these mass shells have all been very close 
together. That is, we assume the big bang to have happened at $t=0$. 
This moment is acquired from the underlying Friedmann model, by 
calculating the age of the Universe $T_0=T_0(H_0,\Omega_0,\Omega_\Lambda)$ 
for the set of cosmological parameters $H_0$, 
$\Omega_0 = \rho_0/\rho_{\rm cr}$, and $\Omega_\Lambda = \rho_\Lambda 
/\rho_{\rm cr}$, where $\rho_{\rm cr}=\frac{3H_0^2}{8\pi G}$, $\rho_0$ is the 
mass density, and $\rho_\Lambda = \frac{\Lambda c^2}{8\pi G}$. 
Then we find the value of the 
energy term $E(r)$ for each present day radius $r_0$, by demanding 
$r(T_0,E)=r_0$ and $r(0,E)=0$,
and integrating backwards in time. The integration is performed numerically
for both  $r$ and $v_r$ with
\begin{equation}
\dot{r}=\sqrt{2\left(\frac{GM(r)}{r}+\frac{\Lambda c^2 r^2}{6} + E\right)}
\label{eq:rdot}
\end{equation}
ensuring the conservation of energy, and
\begin{equation}
\dot{v}_r=-\frac{GM(r)}{r^2}+\frac{\Lambda c^2 r}{3}\: .
\label{eq:vdot}
\end{equation} 
This numerical approach is used because  
for $\Lambda \neq 0$ model a simple parameterized solution given
in e.g.\ Olson \& Silk (\cite{olso79}) does not apply. For $\Lambda=0$,
we checked the numerical calculations by the parameterized formulae. 

For $t\rightarrow 0$,
where the velocities grow considerably, the numerical methods confront some
difficulties. Even with adapted stepsize algorithms we
can not avoid all the problems. Using a fourth order
Runge-Kutta method, we obtain inconsistent results, $\dot{r} \neq v_r$,
when $t<\frac{T_0}{100}$. Fortunately our main concern 
is to calculate the dynamics
at larger $t$, where the method proved to be robust. 
A comparison to the parameterized TB solution showed that the problems 
at small $t$ did not have any effect on our resulting values
of $E(r)$. With the $E(r)$ thus obtained, the present day velocities
$v_r(r_0)|_{E=E(r_0)}$ can be calculated.

\subsection{Testing the method on an N-body simulation \label{sec:sim}}

\begin{figure}
\epsfig{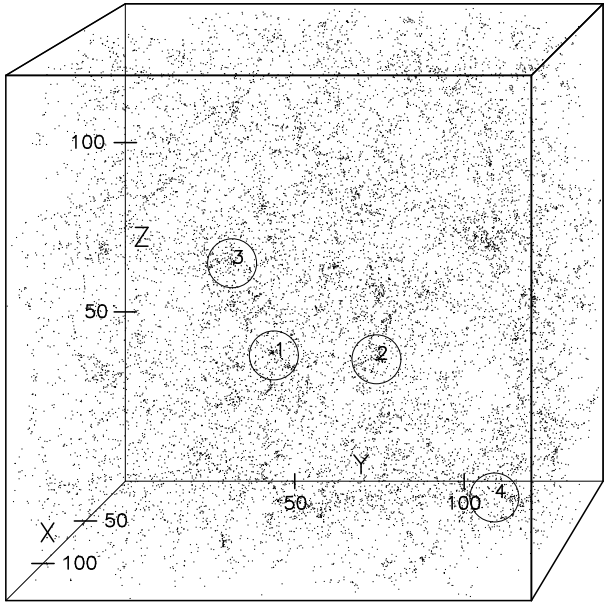}
\caption{The  (141 $h^{-1}$ Mpc)$^3$ cube of GIF N-body simulation galaxies.
The simulation started with 256$^3$ dark halos, producing the 15000 galaxies
seen in this figure. The four circles
indicate the clusters for which we studied the applicability of the 
TB calculations.}
\label{fig:cube}
\end{figure}

\begin{figure}
\epsfig{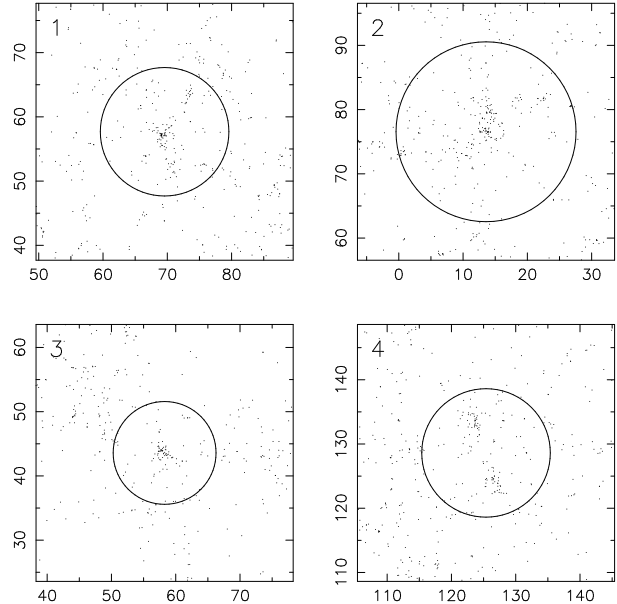}
\caption{The four GIF galaxy clusters extracted for further study. Figures show
XY-planes from the previous figure, centered to the cluster centers. 
The width in Z direction of each plot is 30 Mpc. Circles with
radii $R_{\rm s}$ are drawn. It is notable that these complexes are not
spherically symmetric, resembling the situation with PP.}
\label{fig:sim4c}
\end{figure} 

The TB model is a rough approximation of  the dynamics of galaxy 
clustering. The reader may suspect that the assumption of spherical 
matter distribution  around a galaxy cluster is perilous. In reality
galaxies seem to accumulate into filamentary and 
wall-like structures. However, an averaged galaxy density field may be
well described with a spherical model; the IRAS map, using
a 12 $h^{-1}$ Mpc smoothing, shows a roughly spherical density field 
around PP (Sect.\ \ref{sec:TBap}). And the distribution
of dark matter may be even more spherical. 

How about the peculiar 
velocities? The many-body gravitational interplay determining the
dynamics of galaxy clustering is a complicated issue, can it really
be described by smooth radial TB velocities? That may seem doubtful, but,
again, we rely on the power of averaging. The TB results are compared
only to the average radial velocities around the center of mass.

The correspondence between the TB solutions and the averaged
observational quantities 
is here studied with a cosmological N-body simulation. 
In a simulated galaxy cluster we have the advantage of knowing 
the true distribution of matter
and the true velocities, not distorted by selection effects or observational
uncertainties. And even better, we can study the evolution of the
cluster and compare the matter and velocity distribution of the galaxies
to those predicted by the TB model at different epochs.

In what follows we use the N-body simulation data 
provided by the GIF group, a German-Israeli collaboration
working with the Virgo consortium project ({\tt 
http://star-www.dur.ac.uk/$\sim$frazerp/virgo/}).
The GIF team
investigates the clustering of galaxies in a hierarchical universe
(Kauffmann et al.\, \cite{kauf99a}, \cite{kauf99b}, 
Diaferio et al.\, \cite{diaf99}). They simulate structure formation
in a periodical (141 $h^{-1}$ Mpc)$^3$ cube using
a cold dark matter, flat curvature, non-zero cosmological constant  
model with $\Omega_0=0.3$, $\Omega_\Lambda=0.7$, and $h=0.7$.
The simulation starts with 256$^3$ dark matter halos 
with masses $1.4 h^{-1} 10^{10} M_{\sun}$. The galaxies are formed at 
concentrations of halos, and the effects of gas cooling, 
star formation, and supernova feedback are taken into account.
The data files at 
{\tt http://www.mpa-garching.mpg.de/Virgo/data\_download.html}
have the positions, velocities, and luminosities of galaxies 
and distribution and masses
of halos at six redshifts, $z=$0.00, 0.20, 0.42, 0.62, 0.82,
and 1.05, corresponding to epochs $t/T_0=$ 1.00, 0.97, 0.79, 0.67, 0.57, and
0.49. There are two sets of files, one for the galaxies (luminous matter) and
one for the halos (luminous + dark matter).

Figure \ref{fig:cube}
shows the galaxies in the whole (141 $h^{-1}$ Mpc)$^3$ cube at $z=0$.
We study the evolution of four large clusters in the simulation.
These clusters are circled in Fig.\ \ref{fig:cube} and
shown individually in Fig.\ \ref{fig:sim4c}. For each of these
clusters we approximate the central radius $R_{\rm s}$ by eye,
and calculate the masses and the radially averaged $\delta(r)$,
see Table \ref{tab:sim4c}. These values are similar to what we
observe in the PP supercluster.
\begin{table}
\caption[]{Radii $R_{\rm s}$, and masses and average overdensities
inside $R_{\rm s}$, of the four simulated clusters in Fig.\ \ref{fig:sim4c}.
The units are $[R_{\rm s}]=h^{-1}$ Mpc, $[M]=h^{-1}\,10^{15}M_\odot$.}
\begin{tabular}{l|rrr}
 & $R_{\rm s}$ & $M$ & $\langle \delta \rangle_{R_{\rm s}}$ \\
\hline 1&10&4.5&8.0\\
2&14&7.2&4.3\\
3&8&3.0&10.3\\
4&10&4.4&7.9\\
\end{tabular}
\label{tab:sim4c}
\end{table}

\begin{figure*}
\epsfig{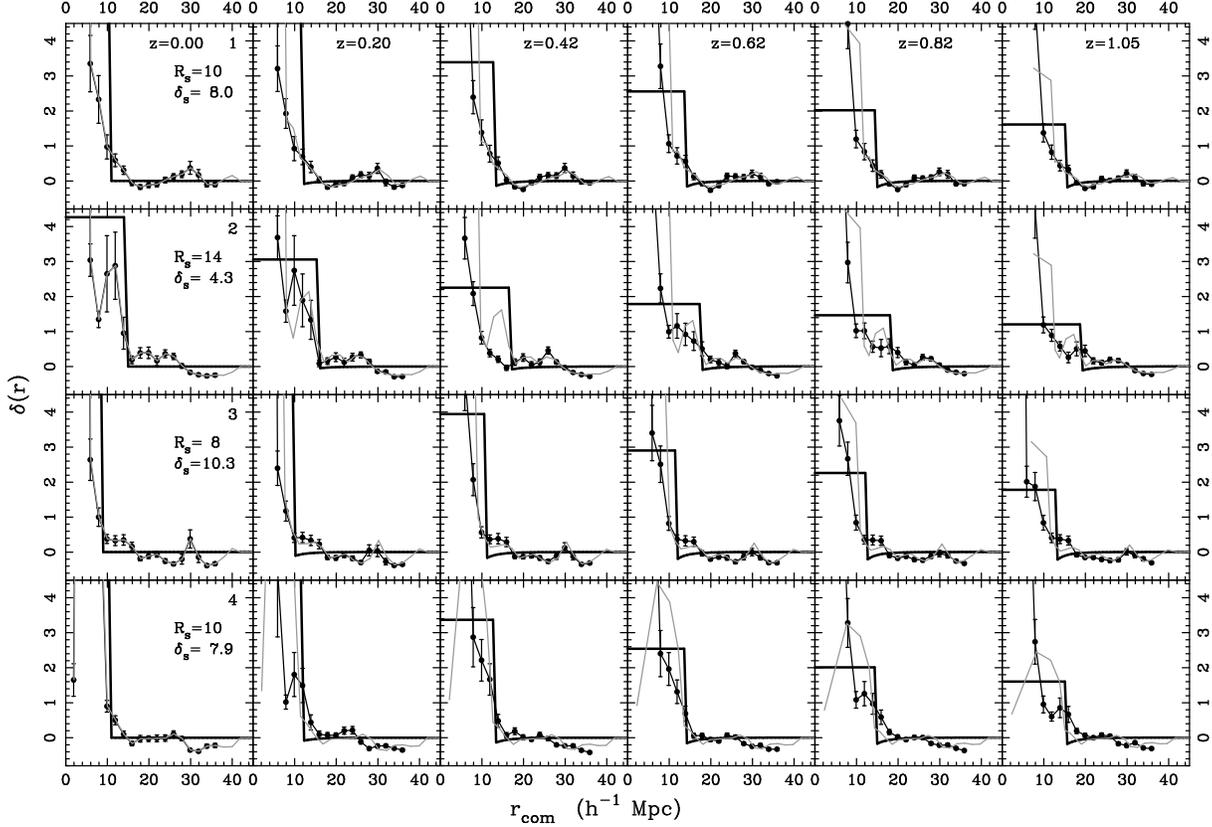}
\caption{The density fields $\delta(r)$ around the four simulated clusters at
the six epochs between the present and $z=1.05$.
Points show the averaged $\delta(r)$ of the simulation, calculated using 
a $3 h^{-1}$ Mpc smoothing. Error bars are the $1\, \sigma$ standard 
deviations. Gray line is the observed density field at $z=0$ evolved
backwards in time via the TB model. Thick black line is the TB evolved
``step function'' toy model. $R_{\rm s}$ are the adopted cluster radii and
$\delta_{\rm s}$ are the average overdensities within these radii at $z=0$.}
\label{fig:simrho}
\end{figure*}
\begin{figure*}
\epsfig{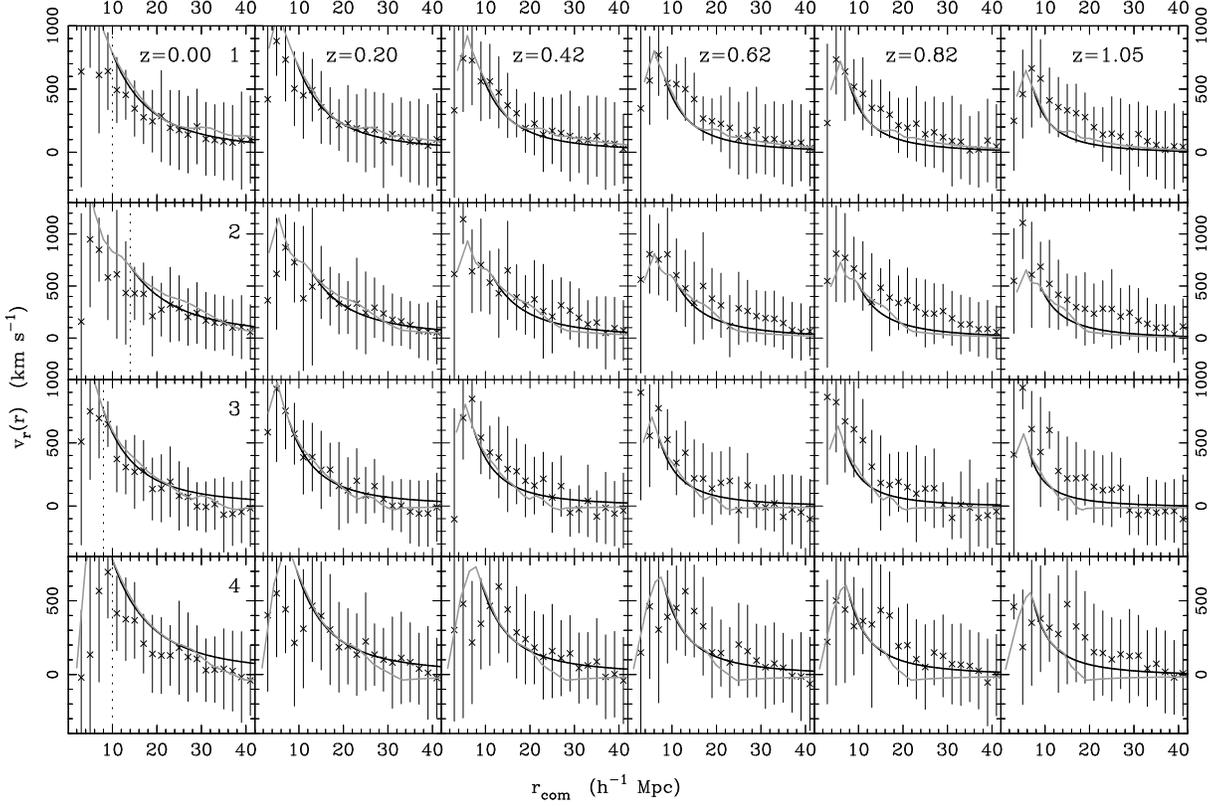}
\caption{The evolution of the velocities $v_r(r)$ around the four clusters.
Crosses and error bars are the averaged $v_r(r)$ and their standard deviations
observed in the simulation. Gray lines correspond to the observed density
distribution and 
black lines to the toy model density distributions, evolved within the
TB models.}
\label{fig:simv}
\end{figure*}

The overdensities are calculated from the 
halo number distribution,  with a $3\, h^{-1}$ Mpc 
smoothing. At this smoothing length 
the matter distribution in the simulation is unbiased according to
Kauffmann et al.\ (\cite{kauf99a}). Then we determine the center of
mass and calculate the radially averaged $\delta(r)$ around each of the
four clusters. For the TB calculations we use either this observed $\delta(r)$
(at $z=0$) or a ``step function'' toy model where $\delta(r< R_{\rm s})=
\langle \delta \rangle_{R_{\rm s}}$, and $\delta(r\geq R_{\rm s})=0$.
Figure \ref{fig:simrho} shows the evolution of $\delta(r)$ in the simulation 
and by means of the TB model. The toy model is very instructive 
in showing how the density enhancements grow in amplitude and converge
in comoving radius with time. 

As expected, in the simulation the structures form at $z>1$ and 
there is not much
evolution in $\delta(r)$ from $z=1.05$ to $z=0$.
Contrarily, in the TB evolved clusters
the overdensities are low at $z=1.05$ but grow quite steadily to the present
day values. At larger $r$ the
$\delta(r)$ given by TB model follow the corresponding values of the simulation
quite well at each epoch.

Figure \ref{fig:simv} shows the comparison of the observed averaged peculiar 
velocities in the simulated clusters to the TB results. A similar time 
dependent factor as in the evolution of the density field is seen here. 
At earlier
epochs the peculiar velocities given by the TB model are somewhat smaller
than the $v_r(r)$ in the simulation. At $z=0$ the TB values are higher,
although the differences are not at a significant level. 

Our conclusion is that concerning the evolution of density and 
velocity fields around clusters there is a marginal difference between
the simulations and the TB calculations.
The overdensities $\delta(r)$ are quite well approximated at large $r$ at
each epoch,
and the present day peculiar velocities $v_r(r)$ are sufficiently 
well derived, that is, with $<1\, \sigma$ deviation, using the
observed present day $\delta(r)$.

\subsection{TB method applied to Perseus-Pisces\label{sec:TBap}}

Now, how to use TB method for Perseus-Pisces where the distribution 
of luminuous matter apparently is not
spherically symmetric? Certain restrictions and assumptions must be applied.
Firstly, we isolate a sphere containing the main part 
of the mass, as was done in Sect.\ \ref{sec:vir}. Then we assume
that any substructures or dynamical features inside the sphere
do not affect a galaxy outside it; the sphere can effectively
be replaced by a point mass located at the center.
Outside the sphere, we assume the density to decrease to the average density
of the universe following spherical symmetry:
\begin{eqnarray*}
\delta(r,\theta,\phi) = \delta(r) \mbox{ if } r>R_{\rm s} \\ 
\delta(r) \rightarrow 0 \mbox{ for } r \gg R_{\rm s}
\end{eqnarray*}
($R_{\rm s} = $ radius of the sphere). After these assumptions we
can apply TB equations for galaxies close to, but not inside the
sphere.

The assumptions may seem doubtful. Is it not oversimplified to
approximate the filamentary structure seen in the PP region with
a spherical model?  The filaments must produce observable non-radial
motions. The main justification for our method lies in the power of 
averaging. Small scale non-radial peculiar velocities are smoothed away
in a similar manner as a smoothed density map blurs away small scale 
substructure. Compare the smoothed IRAS map by Sigad et al.\ (\cite{siga98})
to the galaxy map in Fig.\ \ref{fig:map}. The IRAS map shows a spherical 
density distribution centered in the PP, quite unlike the first glance on
the individual galaxy distribution suggests. N-body simulations indicate
(Sect.\ \ref{sec:sim}) that the same thing happens for the velocity fields,
justifying the usage of the TB models. In any case, over 90 percent of
the matter is dark and may be more symmetrically distributed. The final
criterion is how well the model describes the infall pattern and predicts
the total mass.

\begin{figure}
\epsfig{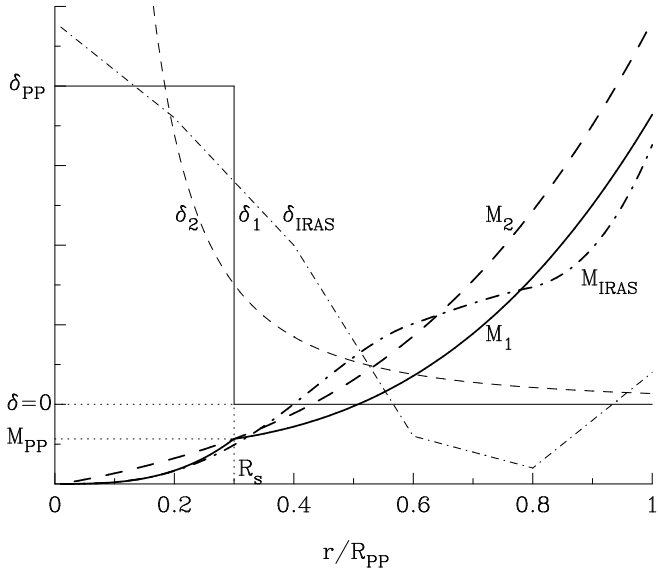}
\caption{Assumed spherically symmetric density distributions 
$\delta_1(r)$, $\delta_2(r)$ and $\delta_{\rm IRAS}(r)$, 
and the corresponding masses inside radius $r$ from the PP center.
Overdensity $\delta_1(r)$ and mass $M_1(r)$ are plotted with continuous,
$\delta_2(r)$ and $M_2(r)$ with dashed, and $\delta_{\rm IRAS}(r)$ and 
$M_{\rm IRAS}(r)$
with dashed-dotted lines. See text for details.}
\label{fig:rho}
\end{figure}

For the density distribution outside the $R_{\rm s}$-sphere 
we consider three models: two toy models, a step function   
\begin{equation}
\delta_1(r)=
\left\{
\begin{array}{ll}
\delta_{\rm PP} & \mbox{ if $r<R_{\rm s}$,} \\ 
0 & \mbox{ if $r \geq R_{\rm s}$,}
\end{array} \right.
\label{eq:rho1}
\end{equation} 
a declining profile
\begin{equation}
\delta_2(r)=Ar^{-2},
\label{eq:rho2}
\end{equation}
where $A$ is chosen so that $M(R_{\rm s})=M_{\rm PP}$,
and a rough approximation of the smoothed density field of IRAS galaxies
around PP, taken from Sigad et al.\
(\cite{siga98}), where the middle panel of their Fig.\ 6 shows
the supergalactic plane with
LG in the center and PP at (X,Y) $\approx$ (50,-10) $h^{-1}$ Mpc. 
It is notable that this is a very rough approximation of the
true matter distribution -- the IRAS map in Sigad et al.\ 
is heavily smoothed ($12 h^{-1}$ Mpc Gaussian filter), and 
the density profile $\delta_{\rm IRAS}(r)$ that we use
is measured by eye from their figure. However, it serves well
for the accuracy that we aim for with the Tolman-Bondi method.
The density models and corresponding cumulative masses are shown
in Fig.\ \ref{fig:rho}.

\begin{figure}
\epsfig{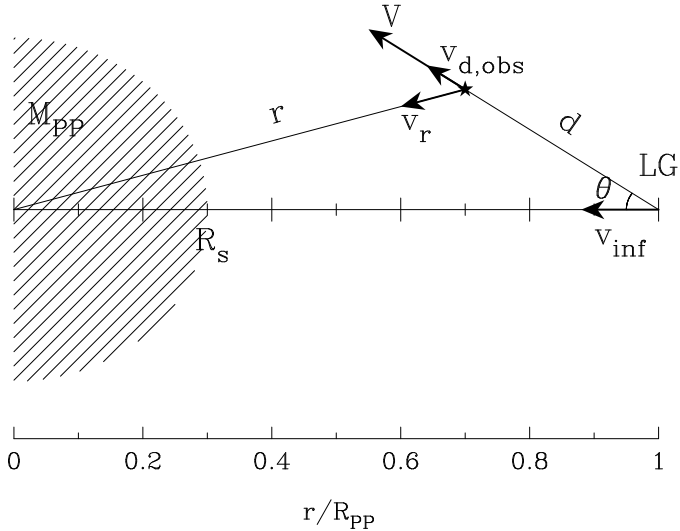}
\caption{Galaxy distances from the PP and LG, and observed and
PP rest frame velocities are plotted in this schematic picture. 
Distances are normalized to the PP distance; $ R_{\rm PP} =1$,
$R_{\rm s}\approx 0.3$. }
\label{fig:sch}
\end{figure}

Figure~\ref{fig:sch} shows schematically the system under study.
A galaxy is located at distance $d$ from us. The lines of sight
towards PP and the galaxy form an angle $\theta$. The distance from PP to the
galaxy is then
\begin{equation}
r=\sqrt{R_{\rm PP}^2+d^2-2 R_{\rm PP} d \cos{\theta}}.
\label{eq:r}
\end{equation}
The observable (radial) component of the peculiar velocity, $v_{d,{\rm obs}}$
(Eq.\ \ref{eq:vpec}),
of the galaxy is connected to the PP velocity field $v_r(r)$ by 
\begin{equation}
v_{d,{\rm obs}}=-v_r(R_{\rm PP})\cos{\theta} \pm v_r(r) 
\sqrt{1-\frac{R_{\rm PP}^2 \sin^2{\theta}}{r^2}},
\label{eq:v_d}
\end{equation}
where $+$ is used for $d<R_{\rm PP} \cos \theta$ and $-$ for 
  $d>R_{\rm PP} \cos \theta$.
If we have a priori knowledge of the LG infall towards PP, 
$v_{\rm inf} \equiv v_r(R_{\rm PP})$
we can solve PP peculiar velocities from observable quantities:  
\begin{equation}
v_r(r)=\pm \frac{v_{d,{\rm obs}}+v_{\rm inf}\cos{\theta}}
{\sqrt{1-R_{\rm PP}^2 \sin^2{\theta}\, /\, r^2}}
\label{eq:v_r}
\end{equation}   
This is the equation used for a comparison of the TB models, giving the
left hand side, and KLUN data, providing the right hand side of the equation.

\section{Comparison of Tolman-Bondi and KLUN sample peculiar velocities
\label{sec:comp}}

\subsection{KLUN-PP sample}

\begin{figure}
\epsfig{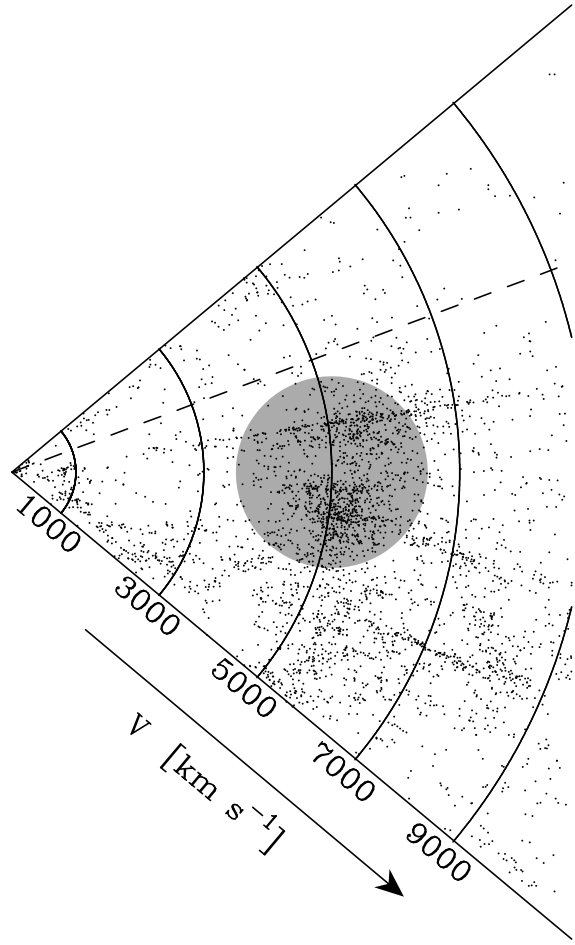}
\caption{LEDA galaxies within angular distance of $40\degr$ from the
PP center. LG is at the apex of the cone, the gray sphere 
indicates the central PP region.
The figure does not give a truthful representation of the galaxy
distribution: Distances based on redshift velocities are distorted,
selection effects are strong in larger distances and projection
effects in the regions close to the cone axis. The figure is a projection
perpendicular to the Galactic plane, which is marked with a dashed
line.}
\label{fig:rmap}
\end{figure}

Peculiar velocities $v_r$ of the KLUN galaxies are obtained in the 
following manner:
\begin{itemize}
\item galaxies are binned into $V$ intervals
and distances $d$ of the bins are calculated
\item each individual galaxy is given the distance of its bin
\item distances from the PP, $r$, and peculiar velocities, $v_r$,
are determined for each galaxy
\item galaxies are binned into $r$ intervals and velocities
$v_r$ of each bin are calculated
\end{itemize}

First we extract a subsample (KLUN-PP) of KLUN galaxies having 
angular distances
from the PP center less than $40\degr$ (Fig.\ \ref{fig:rmap}). 
For these galaxies we calculate 
distances $d$ using the NDM  and the mean surface
brightness dependent diameter TF relation (Sect.\ \ref{sec:ndm}). 
Strictly speaking, the bias correction procedure provides us with an unbiased
average distance at a given $d_{\rm n}$ or, as it can be shown, at a fixed 
kinematical distance $d_{\rm kin}$ or corrected radial velocity. 
Thus we bin the galaxies into 
velocity intervals of $\Delta V = 500$ km~s$^{-1}$, and assign each galaxy 
the distance of its bin. 

To avoid any difficulties in combining possibly incompatible individual
distance measurements we do not use ordinary weighted averages
in the bin distance determinations. The distances are deduced with more
sophisticated Bayesian methods (Press 
\cite{pres97}): Probability distributions of ``good'' and ``bad''
measurements are calculated as
\begin{equation}
P_{{\rm g},i} = \frac{1}{\sqrt{2\pi}\sigma_{\log d,i}}
\exp{\left(- \frac{(\log d_i-\log d_{\rm B}')^2}{2\sigma_{\log d,i}^2} 
\right)}
\label{eq:P_g}
\end{equation}
and
\begin{equation}
P_{{\rm b},i} = \frac{1}{\sqrt{2\pi} S}
\exp{\left(- \frac{(\log d_i-\log d_{\rm B}')^2}{2 S^2} \right)}
\label{eq:P_b}
\end{equation}
where $d_{\rm B}'$ is the (trial) bin distance, and $d_{i}$ are the individual
galaxy TF distances with errors  $\sigma_{\log d,i}$. These individual 
errors are calculated by
\begin{equation}
\sigma_{\log d,i} = \sqrt{\sigma_{\rm TF}^2+\sigma_{\log{D_{25}},i}^2}
\end{equation}
where $\sigma_{\rm TF} = 0.125$  (Theureau \cite{theu98}) 
is the TF relation scatter, and 
$\sigma_{\log{D_{25}}, i}$ are the individual (log) apparent diameter 
measurement errors. In Eq.\ (\ref{eq:P_b}) $S$ characterizes the standard 
deviation of ``wrong'' measurements. In
practice $S$ is assigned a large enough value, exceeding the range where
the measurement error would go unnoticed. We use here 
$S=0.2$ (in units of $\log{d}$).

The probability that the binned set of data ($\mathcal{D}$) have
a distance $d_{\rm B}'$ is given by
\begin{equation}
P(\log d_{\rm B}'|\mathcal{D}) \propto \int_p{\prod_i\left[pP_{{\rm g},i}+
(1-p)P_{{\rm b},i}\right]dp}
\label{eq:P_d}
\end{equation}
following Eq.\ (16) in Press (\cite{pres97}). Here $p$ is the probability 
that individual distance is calculated correctly (and thus $1-p$ is 
the probability for a wrong distance determination). We sum over all 
combinations of individual distance determinations being carried out
correctly/incorrectly \emph{and} over all values of $p$. 
Assuming uniform prior
probability distributions for $p$ and $\log d_{\rm B}'$, the sum reduces to 
Eq.\ (\ref{eq:P_d}).  
The proportionality in Eq.\ (\ref{eq:P_d}) hides the constant needed for 
$\sum_{\log d_{\rm B}'} P(\log d_{\rm B}'|\mathcal{D})=1$. 

\begin{figure}
\epsfig{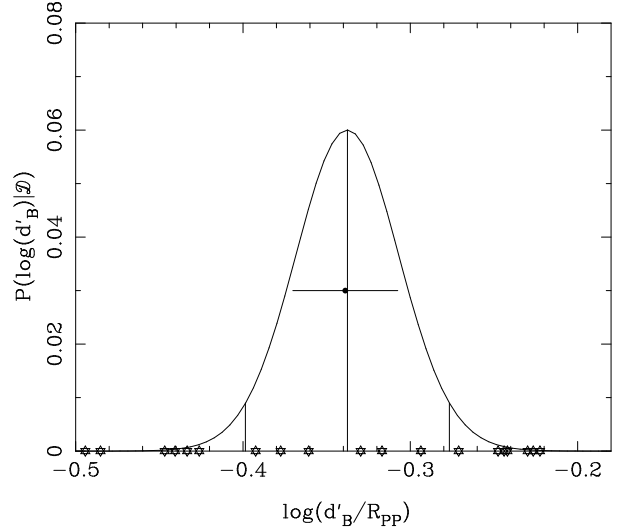}
\caption{An example of using the $P(\log d_{\rm B}'|\mathcal{D})$ 
vs.\ $d_{\rm B}'$
curve in the bin distance determination. Individual galaxies are plotted on
the $d_{\rm B}'$ axis. Vertical lines show the adopted $d_{\rm B}$ and
the 95~\% confidence levels. For comparison, the simple weighted mean
bin distance is also drawn with $2\, \sigma$ error bars. 
KLUN-PP galaxies with 2500~km~s$^{-1}<V<$  
3000~km~s$^{-1}$ are shown in this example.}
\label{fig:P_d}
\end{figure}

\begin{figure}
\epsfig{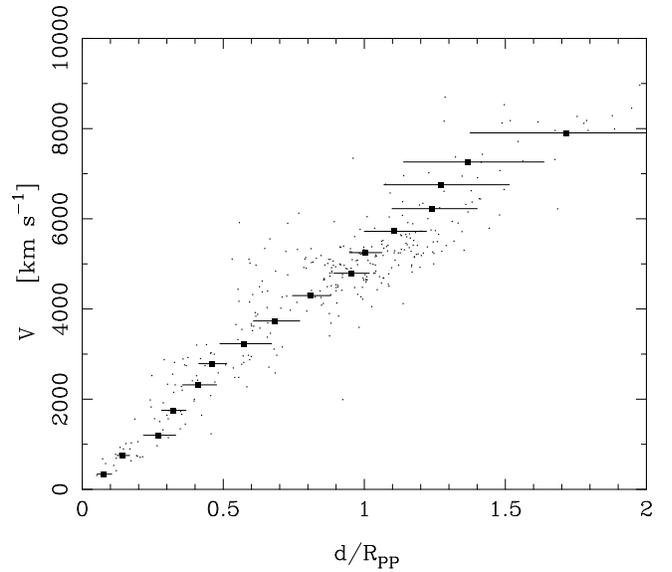}
\caption{KLUN-PP galaxies, bin distances and 95~\% error bars in
a $V$ vs.\ $d$ diagram.}
\label{fig:bin}
\end{figure}

Actual bin distance $d_{\rm B}$ is  adopted  from
the peak of the $P(\log d_{\rm B}'|\mathcal{D})$ vs.\ $\log d_{\rm B}'$ curve,
and 95~\% 
confidence levels are deduced from calculating the area below the curve
(Fig.\ \ref{fig:P_d}). Figure~\ref{fig:bin} 
shows the individual and binned values for all the
data points in the KLUN-PP sample. For comparison the
binned distances are calculated also 
using the simple weighted mean, see Fig.\ \ref{fig:P_d}. 
There are no significant 
differences to the values obtained with the bayesian method, the two
methods deviate $< 1\, \sigma$ for each $d_{\rm B}$.  

The peculiar velocity field around PP is determined similarly. Firstly
each galaxy is assigned its bin distance ($d_{\rm B}$) and 95~\% 
limit errors (denoted by $\sigma_{\rm B}$). 
PP distances ($r_i$) are calculated by Eq.\ (\ref{eq:r}), and errors by
\begin{equation}
\sigma_{r,i}=\max{\left(|r_i-r(d_{\rm B}+\sigma_{\rm B})|,
|r_i-r(d_{\rm B}-\sigma_{\rm B})|\right)}
\label{eq:s_r}
\end{equation}
where $r(x)$ is the function of $d$ given by Eq.\ (\ref{eq:r}). 
Thus Eq.\ (\ref{eq:s_r}) gives
a safe estimate for the maximal error in the 95 \% confidence
interval, unless the galaxy has
\begin{equation}
d_{\rm B}-\sigma_{\rm B} < R_{\rm PP} \cos{\theta} <
d_{\rm B}+\sigma_{\rm B}. 
\label{eq:res}
\end{equation}
The point at $d= R_{\rm PP} \cos{\theta}$ has the lines of sight towards
us and PP exactly perpendicular. There $|r_i-r(d)|$ has a local maximum.
At this point the peculiar
velocity towards PP is also hard to estimate ($\sigma_v$ goes to infinity!).
Thus we reject all galaxies restricted by Eq.\ (\ref{eq:res}). 

Peculiar velocities are obtained with Eq.\ (\ref{eq:v_r}) and
errors by
\begin{equation}
\sigma_{v,i}=\max{\left\{|v_i-v_r(x)| ; x \in (d_{\rm B}-\sigma_{\rm B},
d_{\rm B}+\sigma_{\rm B}) \right\}}
\label{eq:s_v}
\end{equation}
giving again a safe 95 \% confidence level estimate. 
In the end there are 220 galaxies  with
PP-distances and peculiar velocities from 354 galaxies in KLUN-PP before the
restriction of Eq.\ (\ref{eq:res}). From these, 180 have $r>R_{\rm s}$.
Thus roughly half of the original data are lost due to the constraints.

Galaxies are then binned into PP distance intervals of 
$\Delta r = 0.05\, R_{\rm PP}$. For each bin we calculate PP distances and
peculiar velocities with 95\% confidence levels using the Bayesian methods
stated above. Notice that for the peculiar velocity
calculations we need to assume a value for our infall velocity towards PP. 
That is, we have $v_{\rm inf} \equiv v_{r}(R_{\rm PP})$ in 
Eq.\  (\ref{eq:v_r}), as a free parameter.

\subsection{KLUN-PP vs.\ TB}

\begin{figure}
\epsfig{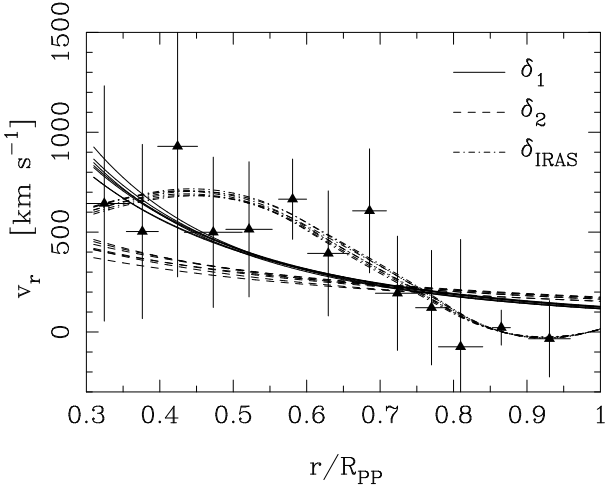}
\caption{Binned KLUN-PP data points and TB $v_r(r)$ curves. KLUN points
are fixed with $v_{\rm inf}=0$. The error bars are $2\, \sigma$
confidence intervals. TB curves are calculated for the three 
density models, $\delta_1$, $\delta_2$, and $\delta_{\rm IRAS}$. For
each density, there is a set of curves, corresponding to the
seven different cosmological models, $(\Omega_0,\Omega_\Lambda )=$ 
(0.1,0.9), (0.1,0.0), (0.2,0.8), (0.2,0.0), (0.4,0.6), (0.4,0.0), and 
(1.0,0.0).}
\label{fig:frv0}
\end{figure}

\begin{figure}
\epsfig{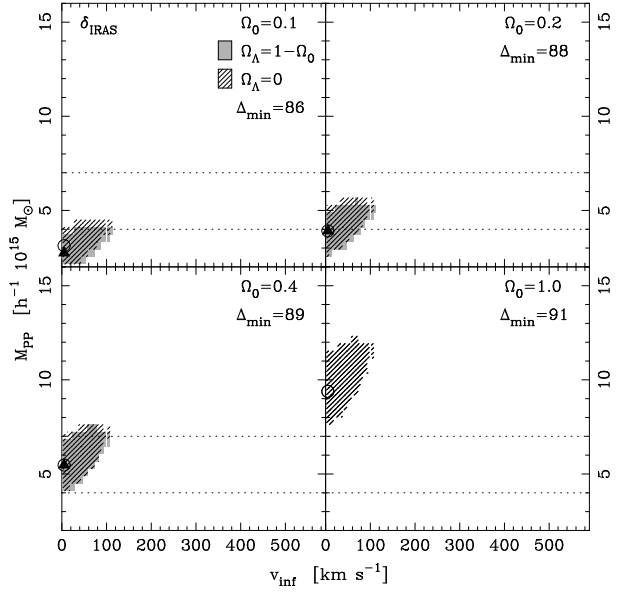}
\caption{The formal $1\, \sigma$ confidence regions (see text) for
$v_{\rm inf}$ and $M_{\rm PP}$. The best fit values are marked with
triangles for $\Omega_\Lambda=1-\Omega_0$ and circles for $\Omega_\Lambda=0$.
 The dotted lines show the $M_{\rm PP}$ limits obtained in Sect.\ 
\ref{sec:vir}.}
\label{fig:fvim}
\end{figure}

We study how well our TB models produce the velocity field of the KLUN galaxies
when the models are varied by
\begin{itemize}
\item  $\Omega_0$, for which we consider values 0.1, 0.2, 0.4, or 1.
Majority of recent studies agrees with this range of values
(Primack \cite{prim00}) 
\item  a choice of either a flat $\Lambda\neq 0$ (i.e.\ $\Omega_\Lambda=
1-\Omega_0$) or an open $\Lambda=0$ universe
\item a choice of density distribution around PP, assumed to be
one of the toy models, $\delta_1$ or $\delta_2$ (Eqs.\ 
\ref{eq:rho1}--\ref{eq:rho2}), or the spherically symmetric approximation
of the IRAS density map, $\delta_{\rm IRAS}$ 
\item mass of the central part of PP, for which we assign values between
2--15 $h^{-1} 10^{15} M_\odot$. In the case of $\delta_{\rm IRAS}$ the mass is
varied by a linear bias factor.
\end{itemize}
KLUN-PP data are adjusted by 
\begin{itemize}
\item LG peculiar velocity towards PP, 
$v_{\rm inf}$ (Eq.\ \ref{eq:v_r}), for which we give values between
0 and 600 km~s$^{-1}$. These limits were chosen because 0 km~s$^{-1}$
is a natural lower limit for a TB application, and after all we expect a
large mass concentration such as the PP supercluster to have at least
some influence on the LG. In the other end, velocities 
exceeding 600 km~s$^{-1}$ would have already been noticed in our motion 
with respect to the CMB, and are therefore not studied.
\end{itemize}

Figure \ref{fig:frv0} shows the comparison of the peculiar
velocities observed for KLUN-PP data and those predicted by the TB models.
Binned KLUN-PP data points with 95 \% confidence level error bars
are plotted in the case of $v_{\rm inf}=0$. The curves are best fit TB
curves to KLUN points, obtained by varying $M_{\rm PP}$ in the
TB calculations. 
Two things are immediately obvious from
looking at Fig.\ \ref{fig:frv0}: Firstly, the IRAS density model fits much
better to the observations than the toy models 
Secondly, different cosmological models give very similar 
best fit curves.

To put this into numbers, we describe the goodness of fit by \[
\chi^2 = \sum_{i=1}^{N}\left(\frac{v_i-\hat{v}_i}{\sigma_i}\right)^2 \]
where $v_i$ is a KLUN data point, $\sigma_i$ its  $1\, \sigma$ deviation,
approximated to be half of the 95 \% error bar, and $\hat{v}_i$ is a
point from a TB curve, at the same $r$ as the KLUN point. 
We get $\chi^2=8.67\pm 0.20$, $12.56 \pm 0.37$, and $2.01 \pm 0.11$, and
significance levels $P=0.796 \pm 0.015$, $0.482 \pm 0.029$, and
$0.99976 \pm 0.00007$
for $\delta_1$, $\delta_2$, and $\delta_{\rm IRAS}$, respectively.
The limits reflect the variance between different cosmologies.

Figure \ref{fig:fvim} shows the formal $1\, \sigma$ confidence regions
for the values of $v_{\rm inf}$ and $M_{\rm PP}$ obtained from fitting
the $\delta_{\rm IRAS}$ TB curves to the KLUN data points. The confidence
regions are defined in the following manner: The value of
$\chi^2$ is calculated for each pair of parameters ($v_{\rm inf}$,
$M_{\rm PP}$). Then a point in the ($v$,$M$) space belongs to the
confidence region if $\chi^2(v,M)<\chi^2_{\rm min}+\tilde{\sigma}_v$,
where $\chi^2_{\rm min}$ is the minimum value of $\chi^2$ in the whole
region of $v=0$--$600$~km~s$^{-1}$ and $M=2$--$15\, h^{-1}\, 10^{15} 
M_\odot$, and $\tilde{\sigma}_v$ is the average of the $1\, \sigma$ errors
of the binned KLUN velocities. This definition thus gives all the points in 
($v_{\rm inf}$,$M_{\rm PP}$) giving $\chi^2 \approx \chi^2_{\rm min}$ 
within observational errors.

\begin{table}
\caption[]{The best fit values and the formal $1\, \sigma$ confidence regions
of $M_{\rm PP}$ and $b_{\rm IRAS}$. See text
for details. The values of $M_{\rm PP}$ are given in units of 
$h^{-1}\, 10^{15}\, M_\odot$.}
\begin{tabular}{ccrr}
$\Omega_0$ & $\Omega_\Lambda$& $M_{\rm PP}$ & $b_{\rm IRAS}$  \\ \hline
  0.1 & 0.9& $ 2.7_{- 0.6}^{+ 1.4}$& $ 0.47_{- 0.14}^{+ 0.08}$ \\
  0.1 & 0.0 & $ 3.1_{- 1.0}^{+ 1.4}$& $ 0.41_{- 0.11}^{+ 0.14}$\\
  0.2 & 0.8 & $ 3.9_{- 1.4}^{+ 1.8}$& $ 0.49_{- 0.15}^{+ 0.24}$\\
  0.2 & 0.0 & $ 3.9_{- 1.4}^{+ 1.8}$& $ 0.49_{- 0.15}^{+ 0.24}$\\
  0.4 & 0.6 & $ 5.5_{- 1.4}^{+ 2.2}$& $ 0.53_{- 0.16}^{+ 0.20}$\\
  0.4 & 0.0& $ 5.5_{- 1.4}^{+ 2.2}$& $ 0.53_{- 0.16}^{+ 0.20}$ \\
  1.0 & 0.0 & $ 9.4_{- 1.8}^{+ 2.9}$& $ 0.55_{- 0.17}^{+ 0.20}$\\
 \end{tabular}
\label{tab:fvim}
\end{table}

The confidence regions in Fig.\  \ref{fig:fvim} suggest that the LG
infall towards PP is less than about 100~km~s$^{-1}$. In each case the
best fit, plotted in the figure as a black triangle for $\Omega_\Lambda=1-
\Omega_0$ and a white circle for $\Omega_\Lambda=0$, is obtained with
$v_{\rm inf}=0$. The best fit values of $M_{\rm PP}$ are coupled
to the value of $\Omega_0$; the correspondence with the range
$M_{\rm PP}= 4$--$7 \, h^{-1}\, 10^{15}\, M_\odot$, which was obtained in
Sect.\ \ref{sec:vir}, is better for $\Omega_0=0.2$--$0.4$ than for
$\Omega_0=0.1$ or $1$.

As mentioned above, the mass $M_{\rm PP}$ is varied by adjusting a
bias parameter for the $\delta$-values given by Sigad et al.\ (\cite{siga98}).
Let $\delta_{\rm S}$ be the overdensity at different distances 
around the PP center, obtained using the contour plot in Fig.\ 6 of
Sigad et al.\ (\cite{siga98}).
Since Sigad et al.\ use $\Omega_0=1$ in
their figure, $\delta_{\rm S}(r)$ must be multiplied by $\Omega_0^{-0.6}$
for a general case of $\Omega_0 \neq 1$. Then we assume a bias of the form
$\Omega_0^{-0.6}\delta_{\rm S}=b_{\rm IRAS}\, \delta_{\rm IRAS}$, 
where $\delta_{\rm IRAS}$ gives the density which we use in 
the TB calculations. 
Increasing $b_{\rm IRAS}$ would mean decreasing the value of $M_{\rm PP}$, 
for example 
when $\Omega_0=0.4$, $M_{\rm PP}=4 \, h^{-1}\, 10^{15}\, M_\odot$
requires $b_{\rm IRAS}=0.83$ and $M_{\rm PP}=7 \, h^{-1}\, 10^{15}\, M_\odot$
means $b_{\rm IRAS}=0.42$. 
The best fit values for $M_{\rm PP}$ and $b_{\rm IRAS}$, 
and their formal $1\, \sigma$ confidence regions, derived from the regions 
in Fig.\ \ref{fig:fvim}, are listed in Table \ref{tab:fvim}.

Our conclusion from comparing the KLUN-PP peculiar velocity
data to the TB-modeled velocity field is that we get a good correspondence
between the two when 
\begin{itemize}
\item the density field around PP is approximated from the IRAS galaxies,
as presented
by Sigad et al.\ (\cite{siga98}), with addition of a linear bias 
$b_{\rm IRAS} \approx 0.5$
\item the infall velocity of the LG towards the center of PP, $v_{\rm inf}$,
is less than $100$ km~s$^{-1}$.
\end{itemize}
We also note that
\begin{itemize}
\item varying the value of $\Omega_0$ between 0.1 and 1 and fixing 
$\Omega_\Lambda =1-\Omega_0$ or 0 have no significant effect on the results,
except that
\item the mass of the central part of the PP agrees with the limits
obtained with the virial theorem if $\Omega_0=0.2$--$0.4$.
\end{itemize}

\begin{figure}
\epsfig{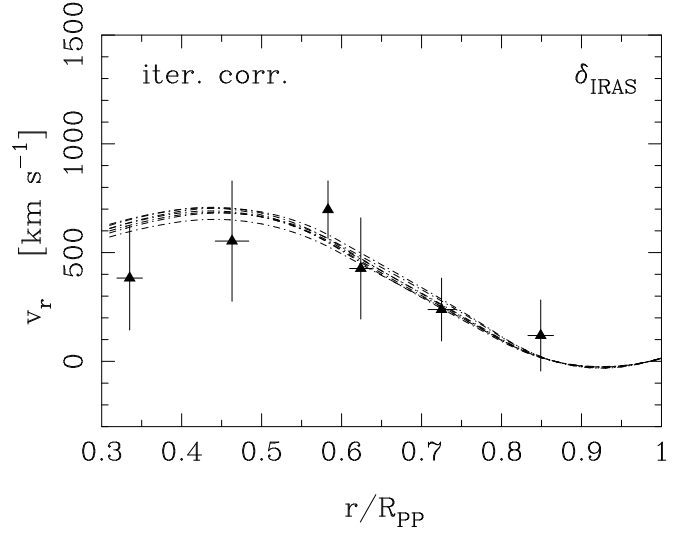}
\caption{Binned data points in $v_r$ vs.\ $r$ diagram after the 
iterative process for correcting distance determinations. Each time
the PP distances are calculated, approximately half of the data are lost
due to restrictions (Eq.\ \ref{eq:res}) and $r>0.3\, R_{\rm PP}$.
Only six binned data points are now present, rather than the 13 shown
in Fig.\ \ref{fig:frv0}. In this figure we have adjusted the KLUN points with
$v_{\rm inf}=0$, and the curves show the best fitted TB solutions using
$\delta_{\rm IRAS}$. The best fit values are $M_{\rm PP}=$
2.9, 3.7, 5.5, and 9.4 $\times h^{-1}\, 10^{15}\, M_\odot$, for $\Omega_0=$
0.1, 0.2, 0.4, and 1.0, respectively, corresponding to bias values 
$b_{\rm IRAS}=$ 0.45, 0.52, 0.53, and 0.55.}
\label{fig:ite}
\end{figure}

\subsection{Iterative corrections, p-class analysis \label{sect:ite}}

The first results can be used to correct
the observed velocities in our NDM distance determination.
With the best fit $\delta_{\rm IRAS}$ TB model shown in Fig.\ \ref{fig:frv0}
we calculate peculiar velocities at different distances $r$, which are used
to  correct the input kinematical distance scale. The whole
process of binning and solving for PP distances and peculiar velocities 
is then repeated with these new kinematical distances as a reference scale. 

A big disadvantage in this iterative procedure is
that at each iteration we lose about half of our data points due to
the constraints of Eq.\ (\ref{eq:res}) and $r> R_{\rm s} = 0.3 \, R_{\rm PP}$.
Figure~\ref{fig:ite} shows how we have only 6 binned data points containing
90 individual galaxies after the first iteration. Our conclusion is that the
iterative process does not in this case improve the results.

Ekholm \& Teerikorpi (\cite{ekho94}) used in their study of the peculiar
velocities in the Great Attractor region a so-called $p$-class analysis.
There the possibility of the $\log d_{\rm TF}$ not being normally
distributed is taken into account by taking only the galaxies lying 
in the unbiased plateau of the  $\langle \log H_0 \rangle(d)$ diagram.
Applying the same procedure here did not give disagreeing results
with  the velocity patterns shown
e.g.\ in Fig.\ \ref{fig:frv0}, but, again, 
the number of points given by the p-class
analysis was small.

\section{Summary\label{sec:res}}

We have studied the velocity field around the 
Perseus-Pisces supercluster using the KLUN galaxy
sample and Tolman-Bondi modeling. In the TB calculations we used a smoothed
density field of IRAS galaxies, as given by Sigad et al.\ (\cite{siga98}).
There is a good correspondence with the peculiar velocities 
given by the KLUN data and the TB models, if the infall
velocity of the LG towards the PP center is assumed to be
less than 100~km~s$^{-1}$ and if we add a bias in
the Sigad et al.\ density field, $b_{\rm IRAS} \approx 0.5$.
The resulting small infall velocity of the LG towards the PP region,
$v_{\rm inf}<100$~km~s$^{-1}$ disagrees with the $v_{\rm inf} \approx
350$~km~s$^{-1}$ of the earlier kinematical studies (see Sect.\ \ref{sec:int}),
but is consistent with the recent POTENT analysis (Dekel \cite{deke99}).   

The goodness of fit between the KLUN and TB velocities is
not affected by varying the cosmological parameters $\Omega_0$
and $\Omega_\Lambda$ within reasonable limits. However, requiring
the mass of the central region of PP to be in the limits
obtained with virial calculations, $M_{\rm PP}=4$--$7\, h^{-1}\,
10^{15}\, M_\odot$, we get constraints for the value of $\Omega_0$. 
Then the intermediate values $\Omega_0=0.2$--$0.4$ are preferred
to the more extreme $\Omega_0=0.1$ or $\Omega_0=1$. 
The value of $\Omega_\Lambda$,
assumed to be either $1-\Omega_0$ or $0$, does not 
have a significant effect on the results. In fact,
the infall velocities around growing structures being little influenced
by the presence of $\Lambda$ is not so unexpected (Peebles \cite{peeb84},
Lahav et al.\ \cite{laha91}).

Calculation of the virial masses of the most prominent clusters in the PP
region resulted in cluster mass to luminosity ratios 
$M/L  = 200$ -- $600 \, h \, M_\odot/L_\odot$. Combining this with 
a luminosity density of the universe $\mathcal{L} = 10^{19.6 \pm 0.1} \,
h$~W~Hz$^{-1}$~Mpc$^{-3}$ (Lilly et al.\ \cite{lill96}), and assuming that
the $M/L$ of the PP region is valid in the rest of the universe, we obtain
$\Omega_0= 0.1$ -- 0.3. Allowing $M/L$ to grow at larger scales would
mean that this is a lower limit. The result agrees with
recent estimates of $\Omega_0$. In comparison, in the study of the Hubble flow
around the Virgo cluster, Teerikorpi et al.\ (\cite{teer92}) found that the 
constraint $M/L=$ constant is inconsistent with $\Omega_0<0.2$ or 
$\Omega_0>1.4$. A smaller $\Omega_0$ was possible if $M/L$ in Virgo
is larger than in the field, contrary to the usual bias scenario. 
The same is valid here.

One of our aims was to study the applicability of the TB method
in a complex environment like the PP supercluster. 
TB solutions are for spherically
symmetric mass distributions only, and looking at the galaxy distribution
in Fig.\ \ref{fig:map}, PP hardly can be taken as such.
However, a smoothed density map, like the IRAS map by Sigad et al.\ 
(\cite{siga98}), has a spherical structure. Similarly, the small scale
dynamics are smoothed away, when we use an averaged velocity map, where
the radial parts of the peculiar velocities are averaged over a shell around
the main galaxy concentration. This is shown not only in the compatibility
of the TB results and the KLUN data, but also in a study of an
N-body simulation. There it was confirmed that the radially averaged
velocity field can be calculated by the TB method, using 
the observed smoothed density
field. There were marginal differences between the
TB solutions and the simulation when we looked at the peculiar velocities
around large clusters at different epochs. However, at the present
epoch these deviations are negligible.
We argue that the Tolman-Bondi method can be used for
a reliable deduction of a radially averaged velocity field around
large galaxy concentrations.

\acknowledgements{We have made use of the Lyon-Meudon Extragalactic Database
(LEDA) supplied by the LEDA team at the CRAL-Observatoire de Lyon (France).
M.H.'s work has been supported by Finnish Cultural Foundation. 
We acknowledge the support by Academy of Finland projects ``Cosmology
in the local galaxy universe'' and ``Galaxy streams and structures in the 
local universe''. The work was initiated while M.H.\ was visiting 
Osservatorio di Capodimonte in Naples. The staff in the observatory was
very kind and warm-hearted and deserves to be mentioned here. We thank Dr.\
Yu.\ Baryshev for his advice concerning the TB model, and the
referee for useful suggestions concerning various parts of the article. }


\begin{thebibliography}{}
\bibitem[1995]{bahc95}
Bahcall, N.A., Lubin, L.M., Dorman, V., 1995, ApJ 447, L81
\bibitem[2000]{bahc00}
Bahcall, N.A., Cen, R., Dav\'{e}, R., Ostriker, J.P., Yu, Q.,
2000, ApJ 541, 1
\bibitem[1947]{bond47}
Bondi, H., 1947, MNRAS 107, 410
\bibitem[1986]{bott86}
Bottinelli, L., Gouguenheim, L., Paturel, G., Teerikorpi, P., 1986, A\&A
156, 157
\bibitem[1992]{bott92}
Bottinelli, L., Durand, N., Fouqu\'{e}, P.\ et al., 1992, A\&AS 93, 173
\bibitem[1993]{bott93}
Bottinelli, L., Durand, N., Fouqu\'{e}, P.\ et al., 1993, A\&AS 102, 57
\bibitem[1995]{bott95}
Bottinelli, L., Gouguenheim, L., Paturel, G., Teerikorpi, P., 1995, A\&A
296, 64
\bibitem[1997]{carl97}
Carlberg, R.G., Yee, H.K.C., Ellingson, E., 1997, ApJ 478, 462
\bibitem[1993]{cour93}
Courteau, S., Faber, S.M., Burstein, D., Willick, J.A., 1993, ApJ 412, L51
\bibitem[1996]{daco96}
da Costa, L.N., Freudling, W., Wegner, G.\ et al., 1996, ApJ 468, L5
\bibitem[1994]{deke94}
Dekel, A., 1994, ARA\&A 32, 371
\bibitem[1999]{deke99}
Dekel, A., Eldar, A., Kolatt, T.\ et al., 1999, ApJ 522, 1 
\bibitem[1996]{dine96}
di Nella, H., Paturel, G., Walsh, A.\ et al., 1996, A\&AS 118, 311
\bibitem[1999]{diaf99}
Diaferio, A., Kauffmann, G., Colberg, J.M., White, S.D.M., MNRAS 307, 552
\bibitem[1996]{ekho96}
Ekholm, T., 1996, A\&A 308, 7
\bibitem[1994]{ekho94}
Ekholm, T., Teerikorpi, P., 1994, A\&A 284, 369
\bibitem[1999]{ekho99}
Ekholm, T., Lanoix, P., Teerikorpi, P., Paturel, G., Fouqu\'{e}, P., 1999,
A\&A 351, 827
\bibitem[2000]{ekho00}
Ekholm, T., Lanoix, P., Teerikorpi, P., Fouqu\'{e}, P., Paturel, G., 2000,
A\&A 355, 835
\bibitem[1995]{freu95}
Freudling, W., da Costa, L.N., Wegner, G.\ et al., 1995, AJ 110, 920
\bibitem[1985]{fouq85}
Fouqu\'{e}, P., Paturel, G., 1985, A\&A 150, 192
\bibitem[1992]{hanm92}
Han, M., Mould, J.R., 1992, ApJ 396, 453
\bibitem[1985]{heis85}
Heisler, J., Tremaine, S., Bahcall, J., 1985, ApJ 298, 8
\bibitem[1997]{huds97}
Hudson, M.J., Lucey, J.R., Smith, R.J., Steel, J., 1997, MNRAS 291, 488 
\bibitem[1993]{jerj93}
Jerjen, H., Tammann, G., 1993, A\&A 276, 1
\bibitem[1999a]{kauf99a}
Kauffmann, G., Colberg, J.M., Diaferio, A., White, S.D.M., 1999a, 
MNRAS 303, 188
\bibitem[1999b]{kauf99b}
Kauffmann, G., Colberg, J.M., Diaferio, A., White, S.D.M., 1999a, 
MNRAS 307, 529
\bibitem[1991]{laha91}
Lahav, O., Lilje, P.B., Primack, J.R., Rees, M.J., 1991, MNRAS 251, 128
\bibitem[1996]{lill96}
Lilly, S.J., Le Fevre, O., Hammer, F., Crampton, D., 1996, ApJ 460, L1
\bibitem[1988]{lynd88}
Lynden-Bell, D., Faber, S.M., Burstein, D.\ et al., 1988, ApJ 326, 19
\bibitem[1999]{mari99}
Marinoni, C., Monaco, P., Giuricin, G., Costantini, B., 1999, ApJ 521, 50
\bibitem[1980]{moul80}
Mould, J., Aaronson, M., Huchra, J., 1980, ApJ 238, 458
\bibitem[1979]{olso79}
Olson, D.W., Silk, J., 1979, ApJ 233, 395
\bibitem[1994]{patu94}
Paturel, G., Bottinelli, L., Gouguenheim, L., 1994, A\&A 286, 768
\bibitem[1997]{patu97}
Paturel, G., Bottinelli, L., di Nella, H.\ et al., 1997, A\&AS 124, 109
\bibitem[1976]{peeb76}
Peebles, P.J.E., 1976, ApJ 205, 318
\bibitem[1984]{peeb84}
Peebles, P.J.E., 1984, ApJ 284, 439
\bibitem[1997]{pres97}
Press, W.H., 1997, Understanding data better with Bayesian and global
statistical methods. In: Unsolved problems in astrophysics, 
Bahcall, J., Ostriker, J.P.\ (eds.),
Princeton University Press (astro-ph/9604126)
\bibitem[2000]{prim00}
Primack, J.R., 2000, Cosmic Flows 1999, ASP Conf.\ Ser.\ vol.\ 201,
Courteau, S., Strauss, M.A., Willick, J.A.\ (eds.) (astro-ph/9912089)
\bibitem[1994]{sand94}
Sandage, A., 1994, ApJ 430, 13
\bibitem[1998]{siga98}
Sigad, Y., Eldar, A., Dekel, A., Strauss, M.A., Yahil, A., 1998, ApJ 495, 516
\bibitem[1998]{smal98}
Small, T.A., Ma, C., Sargent, W.L.W., Hamilton, D., 1998, ApJ 492, 45
\bibitem[2000]{stra00}
Strauss, M.A., 2000, Cosmic Flows 1999, ASP Conf.\ Ser.\ vol.\ 201,
Courteau, S., Strauss, M.A., Willick, J.A.\ (eds.) (astro-ph/9908325)
\bibitem[1975]{teer75}
Teerikorpi, P., 1975, A\&A 45, 117
\bibitem[1984]{teer84}
Teerikorpi, P., 1984, A\&A 141, 407
\bibitem[1997]{teer97}
Teerikorpi, P., 1997, ARA\&A 35, 101
\bibitem[1992]{teer92}
Teerikorpi, P., Bottinelli, L., Gouguenheim, L., Paturel, G., 1992, A\&A
260, 17
\bibitem[1998]{theu98}
Theureau, G., 1998, A\&A 331, 1
\bibitem[1997]{theu97}
Theureau, G., Hanski, M., Ekholm, T.\ et al., 1997, A\&A 322, 730
\bibitem[1998a]{theuA98}
Theureau, G., Rauzy, S., Bottinelli, L., Gouguenheim, L., 1998a, A\&A 340, 21
\bibitem[1998b]{theuB98}
Theureau, G., Bottinelli, L., Coudreau-Durand, N.\ et al., 1998b, A\&AS 
130, 333
\bibitem[1934]{tolm34}
Tolman, R.C., 1934, Proc.\ Nat.\ Acad.\ Sci.\  20, 169
\bibitem[1985]{tull85}
Tully, R., Fouqu\'{e}, P., 1985, ApJS 58, 67
\bibitem[1993]{wegn93}
Wegner, G., Haynes, M.P., Giovanelli, R., 1993, AJ 105, 1251
\bibitem[1990]{will90}
Willick, J.A., 1990, ApJ 351, L5
\bibitem[1991]{will91}
Willick, J.A., 1991, PhD Thesis, University of California, Berkeley
\bibitem[1977]{yahi77}
Yahil, A., Tammann, G., Sandage, A., 1977, ApJ 217, 903
\end{thebibliography}
\end{document}